\documentclass[%
 reprint,
superscriptaddress,
 amsmath,amssymb,
 aps,
 prb,
 floatfix,
]{revtex4-2}

\usepackage{amsmath}
\usepackage{array}
\usepackage{dsfont}
\usepackage{mathtools}
\usepackage{graphicx}
\usepackage{dcolumn}
\usepackage{bm}
\usepackage{hyperref}
\usepackage{physics}
\usepackage{empheq}
\usepackage{color}
\usepackage[x11names]{xcolor}
\usepackage{multirow}
\usepackage{afterpage}
\usepackage{mathrsfs}
\usepackage{bm}
\usepackage{empheq}

\usepackage{tikz}
\usetikzlibrary{shapes}
\usetikzlibrary{tikzmark,calc}

\newcommand{\hexagon}{\mathord{\raisebox{0.6pt}{\tikz{\node[draw,scale=.65,regular polygon, regular polygon sides=6](){};}}}}

\newcommand{\sidc}[2]{\mathbf{#1}_{#2}}

\begin{document}

\title{Finite Temperature Dynamics in 0-Flux and \texorpdfstring{$\bm{\pi}$}{pi}-Flux Quantum Spin Ice:\\
Self-Consistent Exclusive Boson Approach}

\author{F\'elix Desrochers}
\email{felix.desrochers@mail.utoronto.ca}
\affiliation{%
 Department of Physics, University of Toronto, Toronto, Ontario M5S 1A7, Canada
}%
\author{Yong Baek Kim}%
\email{ybkim@physics.utoronto.ca}
\affiliation{%
 Department of Physics, University of Toronto, Toronto, Ontario M5S 1A7, Canada
}%

\date{\today}

\begin{abstract}
Quantum spin ice (QSI) is an emblematic three-dimensional $U(1)$ quantum spin liquid (QSL) on the pyrochlore lattice that hosts gapless photon-like modes and spinon excitations. Despite its notable status and the current rise of strong material candidates Ce$_2$(Zr, Sn, Hf)$_2$O$_7$, there are still only a few analytical approaches to model the low-energy behavior of QSI. These analytical methods are essential to gain insight into the physical interpretation of measurements. We here introduce the self-consistent exclusive boson representation (SCEBR) to model emergent spinon excitations in QSI. By treating the presence of other emergent charges in an average way, the SCEBR extends the range of validity of the exclusive boson representation previously introduced by Hao, Day, and Gingras~\cite{hao2014bosonic} to numerous cases of physical relevance. We extensively benchmark the approach and provide detailed analytical expressions for the spinon dispersion, the Bogoliubov transformation that diagonalizes the system, and the dynamical spin structure factor for 0- and $\pi$-flux QSI. Finite temperature properties are further investigated to highlight essential differences between the thermodynamic behavior of the 0- and $\pi$-flux phases. We notably show that the SCEBR predicts a reduction of the spinon bandwidth with increasing temperature, consistent with previous quantum Monte-Carlo results, through suppression of spinon hopping by thermal occupation. The SCEBR thus provides a powerful analytical tool to interpret experiments on current and future candidate material that has several advantages over other widely used methods.
\end{abstract}
\maketitle


\section{\label{sec: Introduction} Introduction}

\begin{figure}
\centering
\includegraphics[width=1.0\linewidth]{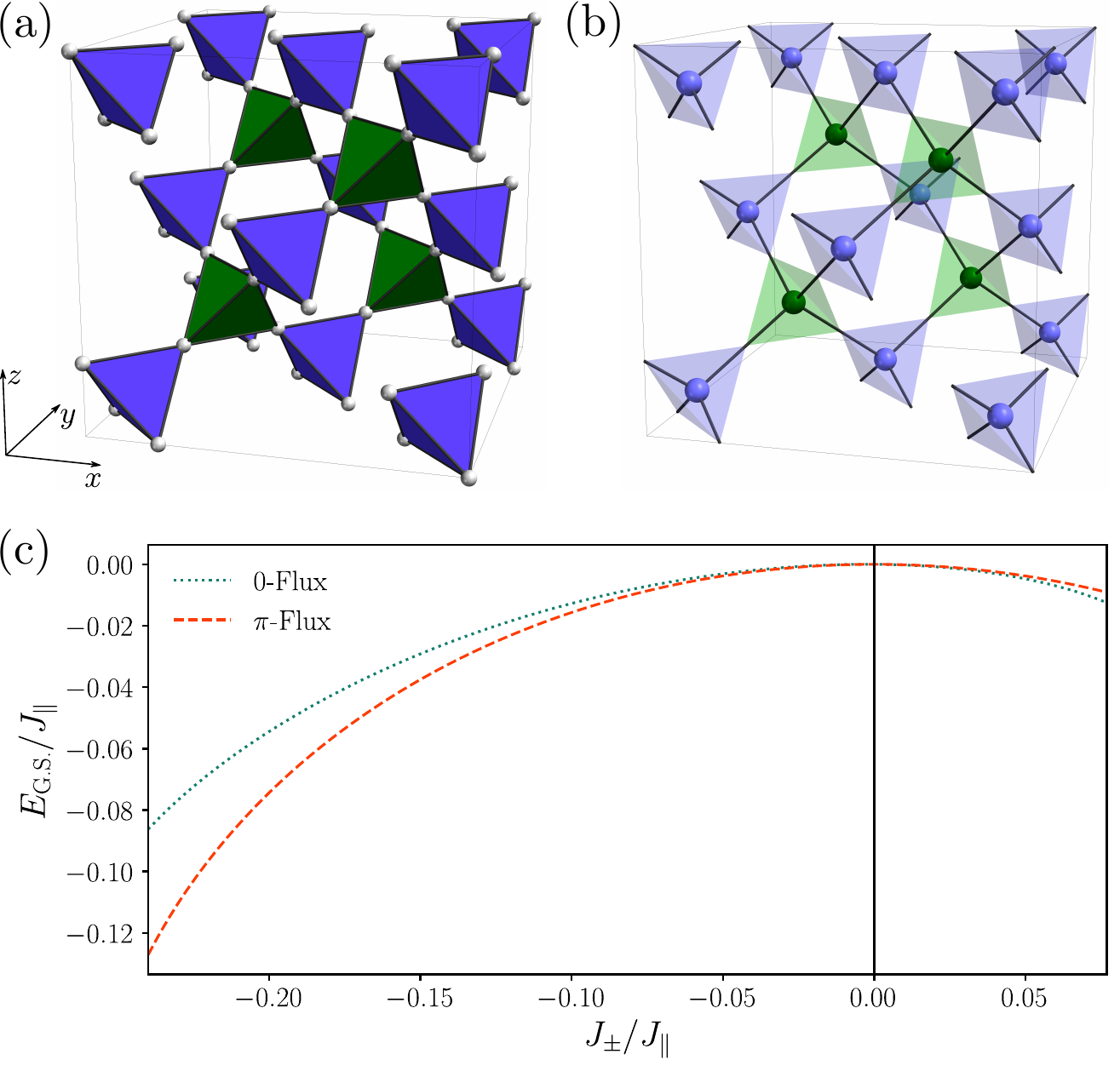}
\caption{(a) The pyrochlore lattice formed by a network of up- (green) and down-pointing (purple) tetrahedra. (b) The pyrochlore lattice sites sit at the bond center of its parent (i.e., premedial) diamond lattice. (c) Ground state energy per diamond lattice unit cell as a function of transverse coupling in the low-density approximation to the exclusive boson formulation for 0- and $\pi$-flux QSI. \label{fig:fig1}}
\end{figure}

A quantum spin liquid (QSL) is a paramagnetic state of a frustrated spin system that supports deconfined fractional excitations and emergent gauge fields as a consequence of its long-range entanglement~\cite{savary2016quantumspinliquids, knolle2019field, zhou2017quantum, balents2010spin, broholm2020quantum, wen2004quantum, wen2013topological, wen2017colloquium, chen2010local}. Quantum spin ice (QSI), which is theoretically predicted to be stabilized on the pyrochlore lattice (see Fig.~\ref{fig:fig1}(a)), is one of the most paradigmatic examples of a QSL. It provides a three-dimensional lattice realization of compact quantum electrodynamics. As such, it hosts emergent gapless photon-like modes as well as gapped spin-1/2 charges of the emergent electric field known as spinons and topological defects (visons)~\cite{hermele2004pyrochlore, benton2012seeing, gingras2014quantum, castelnovo2012spin, udagawa2021spin}. However, the experimental discovery of a QSI has not received definitive evidence and is still a current endeavor in condensed matter physics. 

In that respect, the identification of dipolar-octupolar pyrochlores as promising platforms for QSI~\cite{huang2014quantum} and the subsequent determination that Ce$_2$Sn$_2$O$_7$ is a prospective material realization~\cite{sibille2015candidate} has spawned the beginning of a new era of excitement and breakthrough in the field --- the dawn of new ``Ice Age'' so to speak. The material candidates Ce$_2$Zr$_2$O$_7$~\cite{gaudet2019quantum, gao2019experimental, smith2022case, bhardwaj2022sleuthing, gao2022magnetic, smith2023quantum, Beare2923MuSr}, Ce$_2$Sn$_2$O$_7$~\cite{sibille2015candidate, sibille2020quantum, poree2023fractional}, and Ce$_2$Hf$_2$O$_7$~\cite{Poree2022Crystal, poree2023dipolar} show no sign of ordering or spin freezing down to the lowest accessible temperature. Their microscopic couplings have also been estimated through the fitting of various thermodynamics measurements and determined to be in a region of parameter space that stabilizes the so-called $\pi$-flux QSI where hexagonal plaquettes of the pyrochlore lattice are threaded by a static $\pi$ flux of the emergent gauge field. Some controversy remains about the microscopic parameters of Ce$_2$Sn$_2$O$_7$~\cite{yahne2022dipolar}. Such experimental advances have naturally led to a plethora of theoretical investigations~\cite{li2017symmetry, yao2020pyrochlore, benton2020ground, Placke2020Hierarchy, desrochers2022competing, desrochers2023symmetry, desrochers2023spectroscopic, patri2020theory, Hosoi2022Uncovering, chern2023pseudofermion, yan2023experimentally}. Despite the large number of theoretical works and wide availability of experimental results on strong material candidates, few analytical approaches are available to model the low-energy physics of the $\pi$-flux phase. The only available method is arguably the large-$N$ approximation of gauge mean-field theory (GMFT)~\cite{savary2012coulombic, lee2012generic}. 
Such analytical approaches are highly desirable to understand future and current experiments since they can provide a clear physical interpretation of observed phenomena. They are essential tools to complement brute-force numerical methods. Novel analytical methods have the advantage that they can be comprehensively benchmarked against numerical results~\cite{Hosoi2022Uncovering, chern2023pseudofermion, huang2018dynamics, kato2015numerical} and compared to experiments~\cite{gaudet2019quantum, gao2019experimental, smith2022case, smith2023quantum, poree2023fractional, poree2023dipolar}. 

This work introduces the self-consistent exclusive boson representation (SCEBR) to describe spinon dynamics analytically in QSI. The SCEBR is based on the GMFT approach but does not require the usual large-$N$ approximation. It is instead a direct self-consistent extension of the exclusive boson representation introduced by Hao, Day, and Gingras~\cite{hao2014bosonic}, increasing its range of validity. The extension makes the exclusive boson representation a proper theoretical method to study finite temperature effects and the $\pi$-flux phase in parameter regimes relevant to dipolar-octupolar materials. We first show that the exclusive boson in its initial formulation fails to describe the $\pi$-flux phase beyond the perturbative Ising limit because the low-density approximation of Ref.~\cite{hao2014bosonic} fails. The SCEBR is then introduced to demonstrate that this issue can be mediated by treating the finite density of charges in an average way. In the initial spin model, a finite density of charges inhibits their movement because it is unfavorable for them to hop where a charge is already present since double charge occupancy is energetically suppressed. The essence of the SCEBR is to incorporate this effect by globally reducing the hopping of spinons in a self-consistent manner.

We further explore this construction by extensively studying the XXZ Hamiltonian. Detailed analytical expressions for the spinon dispersions and the Bogoliubov transformations that diagonalize the bosonic quadratic Hamiltonian are provided for the 0- and $\pi$-flux states. These expressions are used to study the dynamic and static properties of QSI. The spinon contribution to the dynamical spin structure factor relevant for experiments on dipolar-octupolar compounds is reported. It is shown that this produces a broad continuum with a single local maximum as a function of energy for 0-flux QSI (0-QSI) and three sharp peaks for $\pi$-flux QSI ($\pi$-QSI). The associated equal-time structure factors show snowflake-like patterns in the $(h,h,l)$ plane with opposite intensity profiles. Such dynamical and static signatures are consistent with 32-site exact diagonalization and previous investigations using the large-$N$ approximation to GMFT~\cite{desrochers2023spectroscopic, Hosoi2022Uncovering}. Finite temperature results are finally explored. Detailed characterization of the thermodynamics behavior highlights that, for 0-QSI, two crossover temperatures are expected to separate the trivial high-temperature paramagnet from the intermediate classical spin ice, and the low-temperature QSI~\cite{banerjee2008unusual, kato2015numerical, huang2018dynamics, huang2020extended}. In contrast, we predict a single crossover directly from the trivial paramagnet to $\pi$-QSI for coupling strengths beyond the perturbative Ising limit. Furthermore, we show that, as the temperature increases, the spinon bandwidth slowly decreases due to a thermal suppression of spinon hopping. Such a thermally induced mobility reduction is consistent with previous quantum Monte Carlo (QMC) investigation~\cite{huang2018dynamics}. The introduced method is one of the first analytical approaches to partially capture such a subtle effect --- a testimony to its potential future usefulness. The SCEBR is thus an analytical approach that has been benchmarked for several use cases and has been shown to be valuable in understanding dipolar-octupolar systems or any future QSI material candidate. 

The rest of the paper is organized as follows: In Sec.~\ref{sec: Theoretical model}, we review the usual GMFT construction and its physical motivation before introducing the SCEBR. We then explore several results obtained with the SCEBR at zero and finite temperature in Sec.~\ref{sec: Results at T=0} and~\ref{sec: Finite temperature results}, respectively. We end by discussing our work's implication and potential future directions in Sec.~\ref{sec: Discussion}.

\section{\label{sec: Theoretical model} Theoretical model}

\subsection{Conventions}

As represented in Fig.~\ref{fig:fig1}(a), the magnetically active ions in spin ice decorate the pyrochlore lattice --- a face-centered cubic Bravais lattice with basis vectors (setting the lattice constant to unity)
\begin{align}
& \mathbf{a}_1=\frac{1}{2}\left(\bf{\hat{y}} + \bf{\hat{z}} \right), \mathbf{a}_2=\frac{1}{2}\left(\bf{\hat{x}} + \bf{\hat{z}} \right), \mathbf{a}_3=\frac{1}{2}\left(\bf{\hat{x}} + \bf{\hat{y}} \right),
\end{align}
and four sublattices. Here $\bf{\hat{x}}$, $\bf{\hat{y}}$, and $\bf{\hat{z}}$ are unit vectors in the global cubic directions $[100]$, $[010]$ and $[001]$. The position of the sublattices can be expressed by $\bm{\delta}_i=\frac{1}{2} \hat{\bf{a}}_i$ $\quad(i=0,1,2,3)$ where we introduced $\hat{\mathbf{a}}_0=\mathbf{0}$ for convenience. Pyrochlore sites are denoted using the following sublattice-indexed pyrochlore coordinates (SIPC)
\begin{align}
& \mathbf{R}_i=\left(r_1, r_2, r_3\right)_{i} = r_1 \mathbf{a}_1+r_2 \mathbf{a}_2+r_3\mathbf{a}_3 + \bm{\delta}_i.
\end{align}

All pyrochlore sites are located at the center of nearest-neighbor bonds on the parent (premedial) diamond lattice illustrated in Fig.~\ref{fig:fig1}(b), whose sites are located at the center of the up- and down-pointing tetrahedra. Each A sublattice (up tetrahedron) of this parent lattice is connected to four B sublattices (down tetrahedra) by the vectors
\begin{align}
& \mathbf{b}_0=\frac{-1}{4}\left(\bf{\hat{x}} + \bf{\hat{y}} + \bf{\hat{z}} \right), \mathbf{b}_1=\frac{1}{4}\left(\bf{-\hat{x}} + \bf{\hat{y}} + \bf{\hat{z}}\right), \nonumber \\
& \mathbf{b}_2=\frac{1}{4}\left( \bf{\hat{x}} - \bf{\hat{y}} + \bf{\hat{z}}\right), \mathbf{b}_3=\frac{1}{4}\left( \bf{\hat{x}} + \bf{\hat{y}} - \bf{\hat{z}} \right).
\end{align}
To label the position of the sites on this parent diamond lattice, we introduce the sublattice-indexed diamond coordinates (SIDC) defined by 
\begin{align}
& \mathbf{r}_\alpha=\left(r_1, r_2, r_3\right)_\alpha=r_1 \mathbf{a}_1+r_2 \mathbf{a}_2+r_3\mathbf{a}_3-\frac{\eta_\alpha}{2} \mathbf{b}_0,
\end{align}
where $\alpha\in\{A,B\}$ labels the diamond sublattice, and $\eta_A=1$ and $\eta_B=-1$ such that $-\eta_\alpha \mathbf{b}_0 / 2$ correspond to the sublattice displacement. 
    
\subsection{Spin Hamiltonian}

To construct the formalism and explore it in the simplest possible setting, we consider the following XXZ model
\begin{align}\label{eq: XXZ model}
    \mathcal{H}=\sum_{\left\langle\mathbf{R}_i \mathbf{R}_j^{\prime}\right\rangle}\left(J_{\parallel} \mathrm{S}_{\mathbf{R}_i}^{\parallel} \mathrm{S}_{\mathbf{R}_j^{\prime}}^{\parallel} - J_{\pm}\left(\mathrm{S}_{\mathbf{R}_i}^{+} \mathrm{S}_{\mathbf{R}_j^{\prime}}^{-}+\mathrm{S}_{\mathbf{R}_i}^{-} \mathrm{S}_{\mathbf{R}_j^{\prime}}^{+}\right)\right),
\end{align}
where $J_{\parallel}>0$, the sum is over nearest-neighbor pyrochlore lattice sites, and the spins are defined in sublattice-dependant local axis coordinates (see Appendix~\ref{Appendix Sec:Conventions}). $\mathrm{S}^+$ and $\mathrm{S}^-$ are raising and lowering operators with respect to $\mathrm{S}^\parallel$ (i.e., $\comm{\mathrm{S}^\parallel}{\mathrm{S}^+}=\mathrm{S}^+$ and $\comm{\mathrm{S}^\parallel}{\mathrm{S}^-}=\mathrm{S}^-$). At this stage, we do not make any assumptions about the transformation properties of the spins. We will only specify whether we are considering effective spin-1/2, non-Kramers doublet, or dipolar-octupolar pseudospins~\cite{rau2019frustrated} when computing physical observables. 

\subsection{Gauge mean-field theory}

\subsubsection{Perturbative regime} \label{subsubsec: perturbative regime}

In the Ising or classical spin ice limit (i.e., $J_{\pm}=0$), the above XXZ model can be rewritten as
\begin{align} \label{eq: CSI model}
    \mathcal{H} = \frac{J_{\parallel}}{2} \sum_{\mathbf{r}_{\alpha}} Q_{\mathbf{r}_\alpha}^2,
\end{align}
where we have defined the charge operator living on the parent lattice 
\begin{align}\label{eq: physical charge GMFT}
    Q_{\mathbf{r}_{\alpha}} = \eta_{\alpha} \sum_{\mu=0}^3 \mathrm{S}^{\parallel}_{\mathbf{r}_{\alpha}+\eta_{\alpha}\mathbf{b}_\mu/2},
\end{align}
and dropped irrelevant constant terms. In this form, it is clear that the energy is minimized by requiring that all tetrahedra have two $\mathrm{S}^{\parallel}$ spins pointing in and two pointing out (i.e., two-in-two-out) such that $Q_{\mathbf{r}_{\alpha}}$ vanishes for all tetrahedra. Since there are many ways to satisfy this energetic constraint known as the ``ice rules," the Ising limit has a degenerate ground state manifold that grows with the system size. 

Upon the addition of a small transverse coupling (i.e., $|J_{\pm}|\ll J_{\parallel}$), one can derive a new effective Hamiltonian within the spin ice manifold by integrating out the gapped charges. The lowest order contribution gives an effective ring-exchange Hamiltonian that can be mapped to a compact $U(1)$ gauge theory of the form~\cite{hermele2004pyrochlore, benton2012seeing}
\begin{align}
    \mathcal{H}_{\text {eff }} \sim-J_{ \pm}^3 / J_{z z}^2 \sum_{\hexagon} \cos (\nabla \times \bar{A})_{\hexagon}.
\end{align}
The sum is over hexagonal plaquettes of the pyrochlore lattice, and the lattice curl is defined by the counterclockwise sum $(\nabla \times \bar{A})_{\hexagon} \equiv \sum_{\langle i, j\rangle \in \hexagon} \bar{A}_{i, j}$. For $J_{ \pm}>0$, it is now well-established from sign-free quantum Monte Carlo (QMC) simulations that the above compact $U(1)$ gauge theory stabilizes a deconfined or Coulomb phase with 0-flux threading the hexagonal plaquettes (i.e., $(\nabla \times \bar{A})_{\hexagon}=0$)~\cite{banerjee2008unusual,benton2012seeing,shannon2012quantum,kato2015numerical,huang2018dynamics, huang2020extended}. This deconfined phase is thus a $U(1)$ QSL, known as 0-flux QSI, that hosts emergent photons as well as gapped monopoles of the emergent electric and magnetic field that are commonly referred to as spinons and visons, respectively~\cite{gingras2014quantum}. In the perturbative treatment outlined above, there exists a unitary transformation between the $J_{\pm}<0$ and $J_{\pm}>0$ regime that maps 0-flux QSI to another QSL known as $\pi$-flux QSI since it is characterized by $(\nabla \times \bar{A})_{\hexagon}=\pi$. However, the sign problem in QMC for $J_{\pm}<0$ makes any theoretical predictions for $\pi$-QSI, apart from its existence close to the Ising point, extremely challenging.

\subsubsection{Slave-spinon construction}

Beyond the perturbative regime, configurations with $Q_{\mathbf{r}_{\alpha}}\ne 0$ will become increasingly important. As such, theoretical description cannot be restricted to the spin ice manifold anymore, and the gapped excitations need to be considered explicitly. Gauge mean-field theory (GMFT), as introduced by Savary and Balents in Ref.~\cite{savary2012coulombic} and further extended in Refs.~\cite{lee2012generic, savary2013spin, savary2016quantum, savary2017disorder, huang2014quantum, li2017symmetry, yao2020pyrochlore, chen2017spectral, yang2021hidden, Yang2020Magnetic, Seth2022Probing, desrochers2023symmetry, desrochers2023spectroscopic}, is a well-established slave-particle construction that describes the $U(1)$ deconfined phases without relying on perturbative arguments. GMFT explicitly considers the gapped matter field by introducing bosonic particles that conceptually correspond to defect tetrahedra breaking the ice rules (i.e., $Q_{\mathbf{r}_\alpha}\ne 0$) at the center of each tetrahedron. To do so, the initial spin-1/2 Hilbert space on the pyrochlore lattice $\mathscr{H}_{\text{spin}}= \otimes_N \mathscr{H}_{\text{S=1/2}}$ is augmented to a new larger one $\mathscr{H}_{\text{big}} = \mathscr{H}_{\text{spin}} \otimes \mathscr{H}_Q$, where $\mathscr{H}_Q$ is the Hilbert space of the bosonic field $Q_{\mathbf{r}_{\alpha}} \in \mathbb{Z}$ that lives on the parent diamond lattice. In this new slave-spinon space, the dominant spin component is mapped to the emergent electric field while the transverse parts are written as spinon bilinears dressed by the emergent photon
\begin{subequations} \label{eq: GMFT mapping for the spins}
\begin{align}
    &\mathrm{S}^+_{\mathbf{r}_{A}+ \mathbf{b}_{\mu}/2} = \frac{1}{2} \Phi^{\dag}_{\mathbf{r}_A}  e^{i A_{\mathbf{r}_{A},\mathbf{r}_{A} + \mathbf{b}_\mu}}  \Phi_{\mathbf{r}_{A}+\mathbf{b}_\mu} \label{eq: GMFT mapping for the spins raising} \\ &\mathrm{S}^{\parallel}_{\mathbf{r}_{\alpha}+\eta_{\alpha}\mathbf{b}_\mu/2} = \eta_{\alpha} E_{\mathbf{r}_{\alpha},\mathbf{r}_{\alpha}+\eta_{\alpha}\mathbf{b}_\mu}. \label{eq: GMFT mapping for the spins Sz}
\end{align}
\end{subequations}
$\Phi_{\mathbf{r}_{\alpha}}^\dag$ and  $\Phi_{\mathbf{r}_{\alpha}}$ are spinon raising and lowering operators that can be written in terms of the canonically conjugate variable to the bosonic charge $\varphi_{\mathbf{r}_{\alpha}}$ (i.e., $\comm{\varphi_{\mathbf{r}_{\alpha}}}{Q_{\mathbf{r}_{\alpha}}}=i$) as $\Phi_{\mathbf{r}_{\alpha}}^\dag = e^{i\varphi_{\mathbf{r}_{\alpha}}}$. These $O(2)$ quantum rotors respect the constraint $|\Phi_{\mathbf{r}_{\alpha}}^{\dagger}\Phi_{\mathbf{r}_{\alpha}}|=1$ by construction. The emergent vector potential $A$ and electric field $E$ are also canonically conjugate and act within the initial $\mathscr{H}_{\text{spin}}$ subspace of $\mathscr{H}_{\text{big}}$. For the above to be a faithful representation of the initial spin Hilbert space, the discretized Gauss's law
\begin{align} \label{eq: constraint gmft gauss's law}
    Q_{\mathbf{r}_\alpha} = \sum_{\mu=0}^3 E_{\mathbf{r}_{\alpha},\mathbf{r}_{\alpha}+\eta_{\alpha}\mathbf{b}_\mu/2} \equiv (\div E)_{\mathbf{r}_{\alpha}}
\end{align}
needs to be enforced on every tetrahedron. Since the spin raising/lowering operator defined in Eq.~\eqref{eq: GMFT mapping for the spins raising} accompanies the flipping of the electric field (by $e^{iA}$) with the creation of a spinon-antispinon pair, it always maps a physical wavefunction to another one that respects Eq.~\eqref{eq: constraint gmft gauss's law}.

Using this construction, the initial XXZ model can be fully rewritten in terms of a compact $U(1)$ lattice gauge theory coupled to quantum rotors
\begin{align} \label{eq: XYZ Hamiltonian with parton operators}
\mathcal{H} =& \frac{J_{\parallel}}{2} \sum_{\mathbf{r}_\alpha} Q_{\mathbf{r}_\alpha}^2  \nonumber \\
&-\frac{J_{\pm}}{4} \sum_{\mathbf{r}_\alpha} \sum_{\mu, \nu \neq \mu} \Phi_{\mathbf{r}_\alpha+\eta_\alpha \mathbf{b}_\mu}^{\dagger} \Phi_{\mathbf{r}_\alpha+\eta_\alpha \mathbf{b}_\nu} \nonumber \\
&\quad\quad\quad\quad\times  e^{i \eta_\alpha\left(A_{\mathbf{r}_\alpha, \mathbf{r}_\alpha+\eta_\alpha \mathbf{b}_\nu}-A_{\mathbf{r}_\alpha, \mathbf{r}_\alpha+\eta_\alpha \mathbf{b}_\mu}\right)}.
\end{align}
The dominant $J_{\parallel}$ term represents the energy cost for the creation of emergent charges, and $J_{\pm}$ leads to an intra-sublattice spinon hopping. Spinons on different diamond sublattices are effectively decoupled. The Hamiltonian further has the following $U(1)$ gauge structure
$$
\begin{aligned}
\Phi_{\mathbf{r}_\alpha} & \rightarrow \Phi_{\mathbf{r}_\alpha} e^{i \chi_{\mathbf{r}_\alpha}} \\
A_{\mathbf{r}_\alpha \mathbf{r}_\beta^{\prime}} & \rightarrow A_{\mathbf{r}_\alpha \mathbf{r}_\beta^{\prime}}-\chi_{\mathbf{r}_\beta^{\prime}}+\chi_{\mathbf{r}_\alpha}
\end{aligned}
$$
as a direct consequence of Gauss's law~\eqref{eq: constraint gmft gauss's law}.

\subsubsection{Mean-field approximation}

To get a tractable model, we fix the gauge field connection to a constant background $A\to\bar{A}$. This effectively decouples the dynamical gauge field $\mathscr{H}_{\text{spin}}$ and matter $\mathscr{H}_{Q}$ sectors. The spinons are then described by coupled quantum rotors with background fields. One can formulate such an approximation as an operator mean-field decoupling~\cite{savary2012coulombic}, a variational calculation~\cite{savary2013spin}, or as a saddle point approximation of a coherent state path integral~\cite{desrochers2023symmetry}. To fix the gauge field background $\bar{A}$, the usual approach is to make an Ansatz with $\nabla\times\bar{A}=0$ for $J_{\pm}>0$ and $\nabla\times\bar{A}=\pi$ for $J_{\pm}<0$ based on the perturbative argument discussed in Sec.~\ref{subsubsec: perturbative regime}. However, as shown in Refs.~\cite{desrochers2023symmetry, desrochers2023spectroscopic}, one can consider symmetry fractionalization in GMFT in a manner analogous to the projective symmetry group classification for Schwinger bosons and Abrikosov fermions parton construction~\cite{wen2002quantum, wang2006spin, messio2013time, bieri2016projective} to compute all gauge field configurations that respect certain symmetries explicitly. Such a treatment shows that the only choices that respect all lattice symmetries (in the absence of external fields) are the 0- and $\pi$-flux states, and thus offer rigorous reasons why we can restrict ourselves to these two Ansätze.

Even after such approximations, one is still left with coupled quantum rotors, which are inherently interacting systems because of the hard constraint on their length. In its most used form, the quantum rotors are then treated using a large-$N$ approximation by relaxing the operator identity $|\Phi_{\mathbf{r}_{\alpha}}^{\dagger}\Phi_{\mathbf{r}_{\alpha}}|=1$ to an average one $\expval{\Phi_{\mathbf{r}_{\alpha}}^{\dagger}\Phi_{\mathbf{r}_{\alpha}}}=\kappa$ which is enforced using a Lagrange multiplier. The most obvious and widely used choice $\kappa=1$ is plagued by many inconsistencies. It produces a dispersion $\varepsilon(\mathbf{k})=J_{\parallel}$ in the Ising limit instead of the expected result $\varepsilon(\mathbf{k})=J_{\parallel}/2$. With $\kappa=1$, GMFT also overestimates the stability of the deconfined phase by predicting that the transition to a magnetically ordered phase occurs at $J_{\pm}/J_{\parallel} \approx 0.192$. In contrast, QMC results show that it should rather appear around $J_{\pm}/J_{\parallel} \approx 0.05$~\cite{banerjee2008unusual, kato2015numerical, shannon2012quantum, huang2018dynamics, huang2020extended}. Because it overestimates the spinon dispersion, GMFT with $\kappa=1$ also predicts that the two-spinon continuum for $0$-flux QSI is about a factor of 2 higher than what is seen in QMC~\cite{huang2018dynamics}. Surprisingly, all of the above-mentioned discrepancies can be cured simultaneously by choosing $\kappa=2$ as explained in Refs.~\cite{desrochers2022competing,desrochers2023symmetry}. In such a case, one recovers the expected dispersion in the Ising limit, agrees with the QMC results of Ref.~\cite{huang2018dynamics} and the exact diagonalization (ED) results of Ref.~\cite{Hosoi2022Uncovering} for the position of the two-spinon continuum in 0- and $\pi$-QSI respectively, and predicts the transition from 0-QSI to the ordered state at the critical value $J_{\pm}/J_{\parallel} \approx 0.048$. Even if physical results are recovered with such a simple change, one could still be worried about the current lack of a clear physical interpretation as to why $\kappa=2$ works and $\kappa=1$ does not. It would thus be advantageous from a physical, analytical, and numerical point of view if one could avoid solving the large-$N$ self-consistency equation.

\subsection{Exclusive boson representation}

The exclusive boson representation of the $O(2)$ quantum rotor introduced by Hao, Day, and Gingras in Ref.~\cite{hao2014bosonic} is an alternative to the above large-$N$ formulation. In such a scheme, the local charge Hilbert space $\mathscr{H}_{Q_{\mathbf{r}_\alpha}}=\left\{ \ket{Q} | Q\in\left\{-2,-1,0,1,2\right\} \right\}$ is split into the product of two bosonic Hilbert spaces $\mathscr{H}_{d}\otimes\mathscr{H}_{b}=\left\{ \ket{n_d,n_b} | n_{d},n_{b}\in \mathds{Z}\right\}$ for the spinons that carry positive ($\mathscr{H}_{d}$) and negative ($\mathscr{H}_{b}$) charges respectively. The initial charge and raising operators are mapped to
\begin{subequations}
\begin{align} \label{eq: initial exclusive boson representation}
    \Phi_{\sidc{r}{\alpha}} &= \frac{1}{\sqrt{1 + d_{\sidc{r}{\alpha}}^{\dagger}d_{\sidc{r}{\alpha}} + b_{\sidc{r}{\alpha}}^{\dagger}b_{\sidc{r}{\alpha}}}} \left( d_{\sidc{r}{\alpha}} + b_{\sidc{r}{\alpha}}^\dagger \right) \\
    Q_{\sidc{r}{\alpha}} &= d_{\sidc{r}{\alpha}}^{\dagger} d_{\sidc{r}{\alpha}} - b_{\sidc{r}{\alpha}}^{\dagger} b_{\sidc{r}{\alpha}},
\end{align}
\end{subequations}
where the creation/annihilation operators satisfy the usual bosonic canonical commutation relations. 

To get a faithful representation of the initial Hilbert space, we enforce the constraint $\left| \Phi^{\dagger}_{\sidc{r}{\alpha}} \Phi_{\sidc{r}{\alpha}}\right|=1$ by requiring that only one species of boson can be on a site at a time. This is equivalent to stating that 
\begin{equation} \label{eq:exclusiveness constraint}
    \left(b_{\sidc{r}{\alpha}}^{\dagger}b_{\sidc{r}{\alpha}}\right) \left(d_{\sidc{r}{\alpha}}^{\dagger}d_{\sidc{r}{\alpha}}\right) = n_{\sidc{r} {\alpha}}^{b} n_{\sidc{r}{\alpha}}^{d} = 0
\end{equation}
for all $\sidc{r}{\alpha}$, which directly implies that 
\begin{equation}
    b_{\sidc{r}{\alpha}} d_{\sidc{r}{\alpha}} = b_{\sidc{r}{\alpha}}^{\dagger}  d_{\sidc{r}{\alpha}}^{\dagger} = 0. 
\end{equation}
In terms of the Hilbert space state, this is equivalent to removing all states of the form $\ket{n_b,n_d}$ where both $n_b\ne 0$ and $n_d\neq 0$, or equivalently by making the unique identification $\ket{Q}=\ket{n_d,0}$ if $Q>0$ and $\ket{Q}=\ket{0,n_b}$ if $Q<0$. This exclusiveness constraint could be imposed by introducing an infinitely large repulsion between the $b$ and $d$ bosons or using a Lagrange multiplier. The finite occupation constraint $|Q_{\mathbf{r}_{\alpha}}|\le 2$ also implies that 
\begin{equation} \label{eq:occupation constraint}
    \left(b^{\dagger}_{\mathbf{r}_\alpha}\right)^{m} = \left(d^{\dagger}_{\mathbf{r}_\alpha}\right)^{m} = b_{\mathbf{r}_\alpha}^{m} = d_{\mathbf{r}_\alpha}^{m} = 0 
\end{equation}
for any $m\ge 3$. 

\subsubsection{Low-density approximation}

To get a tractable model, we will first follow Ref.~\cite{hao2014bosonic} and assume that the density of bosons is small enough that boson-boson interactions, as well as the exclusiveness~\eqref{eq:exclusiveness constraint} and maximum occupation constraints~\eqref{eq:occupation constraint} can be neglected. In doing so, the bosonic charge and raising operators take on the simplified forms
\begin{subequations}
\begin{align}
    \Phi_{\mathbf{r}_{\alpha}} &\approx d_{\mathbf{r}_{\alpha}} + b_{\mathbf{r}_{\alpha}}^\dagger\\
    Q_{\mathbf{r}_{\alpha}}^2 &\approx d_{\mathbf{r}_\alpha}^\dagger d_{\mathbf{r}_\alpha} + b_{\mathbf{r}_\alpha}^\dagger b_{\mathbf{r}_\alpha}. 
\end{align}
\end{subequations}
The Hamiltonian, for the matter sector, is then
\begin{widetext}
\begin{align} 
\mathcal{H} =&\frac{J_{\parallel}}{2} \sum_{\mathbf{r}_\alpha} (d^{\dagger}_{\mathbf{r}_\alpha}d_{\mathbf{r}_\alpha} + b^{\dagger}_{\mathbf{r}_\alpha}b_{\mathbf{r}_\alpha} )  \nonumber \\
&-\frac{J_{\pm}}{4} \sum_{\mathbf{r}_\alpha} \sum_{\mu, \nu \neq \mu} e^{i \eta_\alpha\left(\bar{A}_{\mathbf{r}_\alpha, \mathbf{r}_\alpha+\eta_\alpha \mathbf{b}_\nu}-\bar{A}_{\mathbf{r}_\alpha, \mathbf{r}_\alpha+\eta_\alpha \mathbf{b}_\mu}\right)}\left( d_{\mathbf{r}_\alpha+\eta_\alpha \mathbf{b}_\mu}^{\dagger} + b_{\mathbf{r}_\alpha+\eta_\alpha \mathbf{b}_\mu} \right)  \left( d_{\mathbf{r}_\alpha+\eta_\alpha \mathbf{b}_\nu} + b_{\mathbf{r}_\alpha+\eta_\alpha \mathbf{b}_\nu}^{\dagger} \right),
\end{align}
\end{widetext}
where the gauge field background is chosen such that $\curl{\bar{A}}=0$ and $\curl{\bar{A}}=\pi$ for the 0-flux and $\pi$-flux Ansätze, respectively. Many different gauge-equivalent choices yield the same fluxes and physical results. In the following, we use the gauge fixing of Ref.~\cite{desrochers2023symmetry} summarized in Appendix~\ref{Appendix Sec:Analytical diagonalization of the quadratic Hamiltonians}. The above low-density approximation to the exclusive boson representation has several advantages over the large-$N$ approach. First, it disentangles all physical processes hidden in the hopping term $\Phi^\dagger \Phi$ (i.e., spinon-antispinon pair creation/annihilation and positive/negative spinon hopping) as shown in the Hamiltonian above. Second, there are no self-consistency conditions to solve. Last, the physical justification for the approximation and its expected regime of validity are clear. Indeed, the above approximation is evidently valid close to the Ising limit ($|J_{\pm}|\ll J_{\parallel}$) at low temperature ($T\ll J_{\parallel}$). Beyond such a regime, the validity of the low-density approximation is uncertain and needs to be verified. 

\begin{figure}
\centering
\includegraphics[width=1.0\linewidth]{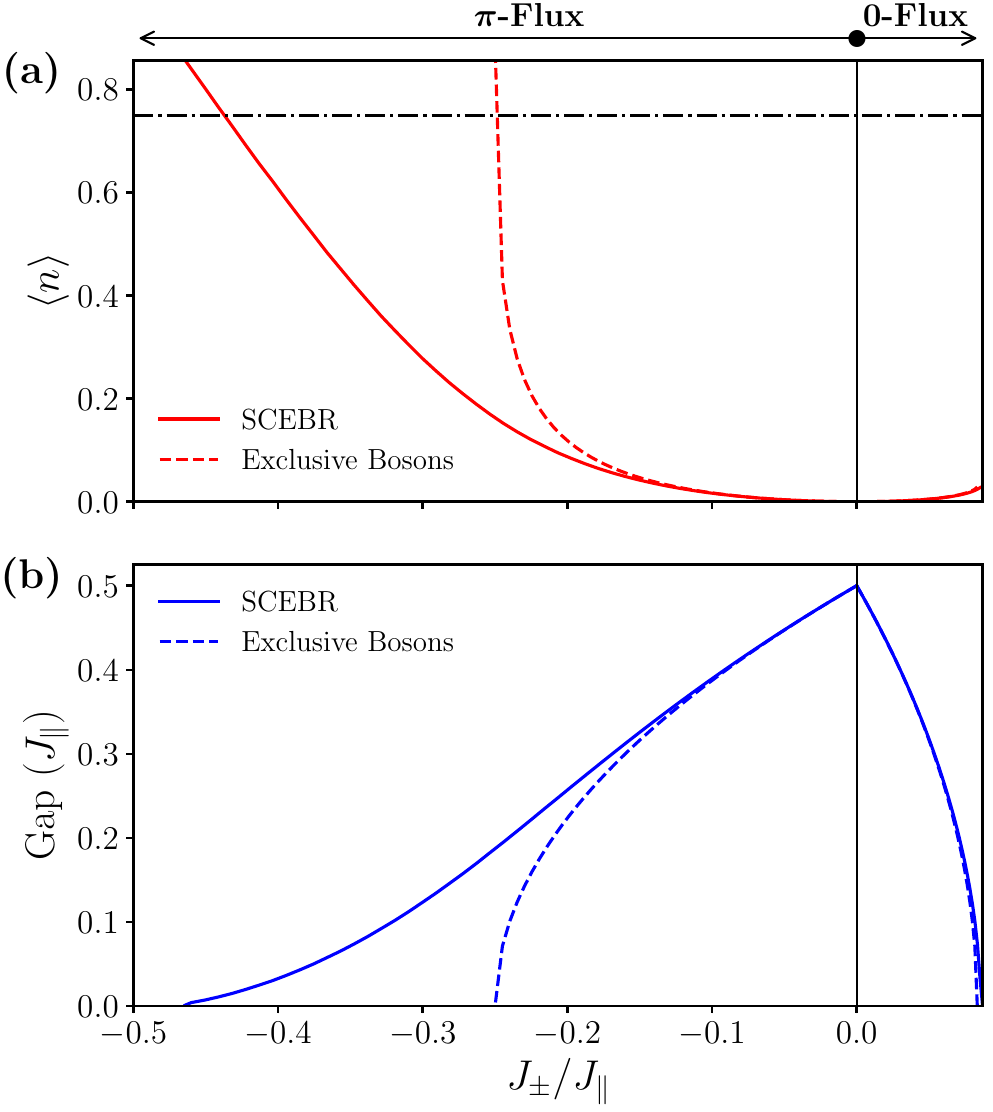}
\caption{(a) Average tetrahedron occupation and (b) spinon gap as a function $J_{\pm}/J_{\parallel}$ using the low-density exclusive boson approach (dashed lines) and the self-consistent exclusive boson representation (full lines). Calculations use 0-flux and $\pi$-flux QSI for $J_{\pm}>0$ and $J_{\pm}<0$, respectively (i.e., the lowest energy state). \label{fig:density and gap SCEBR}}
\end{figure}

A byproduct of the absence of self-consistency conditions to solve is that the above boson Hamiltonian can be diagonalized analytically. We present in Appendix~\ref{Appendix Sec:Analytical diagonalization of the quadratic Hamiltonians} the analytical form of the dispersion and the Bogoliubov transformation that diagonalize the system~\cite{colpa1978diagonalization} for 0- and $\pi$-QSI. The resulting ground state energy per diamond lattice unit cell as a function of $J_{\pm}$ is shown in Fig.~\ref{fig:fig1}(c). The 0-flux and $\pi$-flux phases are more stable for $J_{\pm}>0$ and $J_{\pm}<0$, respectively, as expected from the above perturbative argument. We also present in Fig.~\ref{fig:density and gap SCEBR} the spinon dispersion gap and tetrahedron occupation density 
\begin{equation}\label{eq: average tetrahedron occupation}
    \expval{n_{\alpha}} = \frac{1}{N_{\text{d.u.c.}}} \sum_{\mathbf{r}} \left(\expval{d_{\mathbf{r}_\alpha}^{\dagger}d_{\mathbf{r}_\alpha}} + \expval{b_{\mathbf{r}_\alpha}^{\dagger}b_{\mathbf{r}_\alpha}}\right),
\end{equation}
where $N_{\text{d.u.c.}}$ is the number of diamond lattice unit cells and $\expval{n}=\expval{n_A}=\expval{n_B}$ because of the sublattice symmetry. We see that the bosons condense (i.e., the spinon gap vanishes) at $J_{\pm}/J_{\parallel}=1/12$ for the 0-flux state and at $J_{\pm}/J_{\parallel}=-1/4$ for the $\pi$-flux state. When the bosons condense, the $U(1)$ QSL undergoes an Anderson-Higgs transition where the $U(1)$ gauge fluctuations are gapped out. This condensation point thus signals a transition to a magnetically long-range ordered phase.

We here comment on a subtle but important point. It is often mentioned that the spinons are conceptually analogous to the tetrahedra that break the ice rules. One might then be tempted to identify $\expval{n}$ with the spinon density. However, the above conceptual intuition is only approximately valid close to the Ising limit. Indeed, at zero temperature, the ground state is always a spinon vacuum (see Appendix~\ref{Appendix Sec:Analytical diagonalization of the quadratic Hamiltonians} for details). The true spinons are quantum coherent excitations above the QSI ground state, which is formed by a large superposition of quantum states. In the perturbative regime, this superposition is mostly formed by states in the 2-in-2-out manifold, implying that $\expval{n}\approx 0$. However, far from the Ising point, spin configurations that break the ice rules can also have a significant contribution (i.e., $\expval{n}\ne 0$). We then emphasize that $\expval{n}$ is not to be confused with the average spinon number. As such, we will refer to it as the average tetrahedron occupation in the rest of the article. 

Coming back to Fig.~\ref{fig:density and gap SCEBR}, the tetrahedron occupation remains negligible even at the transition point for 0-QSI. The gap also closes for a similar value to QMC. This justifies \emph{a posteriori} the small density approximation for the 0-flux phase. In contrast, for the $\pi$-flux phase, the transition seen around $J_{\pm}/J_{\parallel}=-1/4$ contradicts numerous existing results. Indeed, the large-$N$ approach to GMFT~\cite{lee2012generic, desrochers2023spectroscopic}, ED~\cite{patri2020distinguishing, benton2020ground}, pseudo-fermion renormalization group~\cite{chern2023pseudofermion}, finite temperature Monte Carlo~\cite{taillefumier2017competing}, numerical linked cluster, series expansion~\cite{benton2018quantum}, and convolutional neural network quantum states variational calculations~\cite{astrakhantsev2021broken} all predict that there should not be any phase transition as $J_{\pm}/J_{\parallel}$ is tuned from the Ising to the Heisenberg point (i.e., $J_{\pm}/J_{\parallel}=-1/2$). The above failure of the low-density approximation to the exclusive boson representation can be rationalized by noticing the sharp rise of the density around the condensation point in Fig.~\ref{fig:density and gap SCEBR}(a). The tetrahedron occupation even rises above the infinite temperature limit $\expval{n}=3/4$. When $T\to\infty$ all 16 tetrahedron configurations become equally likely (i.e., six 2-in-2-out, four 3-in-1-out, four 1-in-3-out, and two all-in-all-out) such that $\expval{n}=3/4$. In this high-occupation regime, the low-density approximation fails and is naturally expected to yield unphysical results.

\subsubsection{Self-consistent scheme}

Although the above ``zeroth-order" low-density approximation is valid for small values of the transverse coupling $|J_{\pm}|$ and low temperatures, we have seen that it breaks down in physically relevant cases where the density of spinons becomes non-negligible, such as the $\pi$-QSI for large $|J_{\pm}|$. To improve the quantitative agreement of the theory, a natural approach would be to try and rigorously consider bosons-boson interactions generated by higher-order terms in the density expansion and the exclusiveness constraint. 

Instead, we propose to consider the spinon interactions in a mean-field manner by introducing the \emph{self-consistent exclusive boson representation} (SCEBR) of the quantum rotor. In this SCEBR, the annihilation operator of Eq.~\eqref{eq: initial exclusive boson representation} is replaced by
\begin{align}
    \Phi_{\sidc{r}{\alpha}} &\approx \frac{1}{\sqrt{1 + \expval{n_{\alpha}}}} \left( d_{\sidc{r}{\alpha}} + b_{\sidc{r}{\alpha}}^\dagger \right),
\end{align}
where the mean-field parameter $\expval{n_\alpha}$ is determined self-consistently using Eq.~\eqref{eq: average tetrahedron occupation}. In this framework, the spinon hopping term is now
\begin{align}
    &J_{\pm} \Phi_{\mathbf{r}_{\alpha}+\eta_{\alpha}\mathbf{b}_{\mu}}^{\dagger} \Phi_{\mathbf{r}_{\alpha}+\eta_{\alpha}\mathbf{b}_{\nu}} \nonumber \\
    &\approx \frac{J_{\pm}}{1 + \expval{n_\beta}} \left( d_{\mathbf{r}_\alpha+\eta_\alpha \mathbf{b}_\mu}^{\dagger} + b_{\mathbf{r}_\alpha+\eta_\alpha \mathbf{b}_\mu} \right) \nonumber \\
    &\hspace{2cm}\times\left( d_{\mathbf{r}_\alpha+\eta_\alpha \mathbf{b}_\nu} + b_{\mathbf{r}_\alpha+\eta_\alpha \mathbf{b}_\nu}^{\dagger} \right),
\end{align}
where $\beta\ne\alpha$. It can then be seen that the only effect of the SCEBR is to renormalize the transverse coupling $J_{\pm}$ to the smaller value
\begin{equation}\label{eq: renormalized transverse projector}
    \tilde{J}_{\pm} = \frac{J_{\pm}}{1 + \expval{n}}, 
\end{equation}
where once again $\expval{n}=\expval{n_A}=\expval{n_B}$ because the two sublattices are decoupled. Therefore, the presence of tetrahedra that break ice rules inhibits, in an average way, the hopping of spinons on the lattice.

Looking at  Fig.~\ref{fig:density and gap SCEBR}, the results for 0-QSI using the SCEBR are, for all intent and purposes, unaffected since the tetrahedron occupation always remains negligible such that $\tilde{J}_{\pm}\approx J_{\pm}$. For $\pi$-QSI, the position of the magnetic transition significantly changes and occurs just before the Heisenberg point. This prediction is in much better agreement with existing results from the literature. Hence, the SCEBR is expected to be valid for a larger parameter regime than the low-density approximation. It thus makes the exclusive boson representation a useful analytical tool in more physical situations of interest. The following sections explore physical predictions from the SCEBR at zero and finite temperatures.

\section{\label{sec: Results at T=0} Results at \texorpdfstring{$\bm{T=0}$}{T-0}}

We first explore physical predictions of the SCEBR at $T=0$ for both the 0- and $\pi$-flux phases. The main purpose of this section is to benchmark the $T=0$ results of the SCEBR in comparison to other methods such as the large-$N$ approach~\cite{desrochers2023spectroscopic} and exact diagonalization~\cite{Hosoi2022Uncovering}. We find that the agreement between the SCEBR with previous theoretical investigations and experiments~\cite{poree2023fractional, gaudet2019quantum, smith2022case} gives strong support to its validity. In the next section, we move on to use the SCEBR in a relatively less explored theoretical territory: spinon behavior at finite temperatures.  

Notice that the only effect of the SCEBR, in comparison to the previous exclusive boson approach without self-consistency, is to replace the microscopic transverse coupling $J_{\pm}$ by the renormalized one $\tilde{J}_{\pm}$. Therefore, the analytical expression for the dispersion and Bogoliubov transformation that diagonalize the system with the SCEBR are the same as in the low-density approximation with the simple replacement $J_{\pm}\to \tilde{J}_{\pm}$. Detailed derivations of these expressions are given in Appendix~\ref{Appendix Sec:Analytical diagonalization of the quadratic Hamiltonians}. For 0-QSI, there is a single spinon band of the form 
\begin{align}
    \varepsilon^{\text{0-flux}}(\mathbf{k}) &= \frac{1}{2} \sqrt{ J_{\parallel} \left( J_{\parallel} - 2\tilde{J}_{\pm}\sum_{a,b\ne a}\cos\left(\frac{k_a}{2}\right) \cos\left(\frac{k_b}{2}\right) \right)},
\end{align}
where the wavevector is written in global Cartesian coordinates $\mathbf{k}=k_x \mathbf{\hat{x}} + k_y\mathbf{\hat{y}} + k_z\mathbf{\hat{z}}$. 
It is then straightforward to see that the spinon gap is located at the zone center and given by
\begin{equation}
    \Delta_{\text{spinon}}^{\text{0-flux}} = \frac{1}{2} \sqrt{J_{\parallel} \left( J_{\parallel} - 12 \tilde{J}_{\pm}\right)},
\end{equation}
such that the critical value of the transverse coupling at which the spinons condense (i.e., the transition point from the $U(1)$ QSL to the ordered state) is
\begin{equation}
    \left( \frac{\tilde{J}_{\pm}}{J_{\parallel}} \right)_{c}^{\text{0-flux}} = \frac{1}{12}
\end{equation}
as mentioned previously. 

For $\pi$-QSI, translation symmetry is fractionnalized~\cite{essin2013classifying, essin2014spectroscopic, chen2017spectral, desrochers2023symmetry, desrochers2023spectroscopic}. This leads to an enlargement of the unit cell and a spectral periodicity enhancement of the dispersion. We then find that $\pi$-QSI has two non-degenerate bands 
\begin{widetext}
\begin{align}
    \varepsilon_{\pm}^{\pi\text{-flux}}(\mathbf{k}) &= \frac{1}{2} \sqrt{J_{\parallel} \left( J_{\parallel} \pm 2 \sqrt{ \tilde{J}_{\pm}^2 \left( 3 - \sin\left(k_x\right) \sin\left(k_y\right) - \sin\left(k_x\right)\sin\left(k_z\right) - \sin\left(k_y\right)\cos\left(k_z\right)   \right)} \right)} 
\end{align}
\end{widetext}
with a gap of
\begin{equation}
    \Delta_{\text{spinon}}^{\pi\text{-flux}} = \frac{1}{2} \sqrt{J_{\parallel}\left(J_{\parallel} - 4 |\tilde{J}_{\pm}|\right)},
\end{equation}
which implies that the spinons condense at 
\begin{equation}
    \left(\frac{\tilde{J}_{\pm}}{J_{\parallel}}\right)_c^{\pi\text{-flux}}=\frac{1}{4}.
\end{equation}
The above dispersion has a particular structure, which is also present in the large-$N$ approach~\cite{lee2012generic, desrochers2023spectroscopic}. There are lines of dispersion minima (maxima) that occur for the $\varepsilon_{-}^{\pi\text{-flux}}(\mathbf{k})$ ($\varepsilon_{+}^{\pi\text{-flux}}(\mathbf{k})$) band at momenta
\begin{align}
    k_a=\left(n+\frac{1}{2}\right) \pi, \hspace{1.5mm} k_b=\left(n+ 2m -\frac{1}{2}\right) \pi, \hspace{1.5mm} k_c \in \mathbb{R},
\end{align}
where $n,m \in \mathbb{Z}$, and $(a, b, c)$ is any permutation of the $(x, y, z)$ triplet.

Next, to make a connection with experiments and compare the SCEBR with previous theoretical investigations using the large-$N$ theory, we compute the dynamical spin structure factor 
\begin{align}\label{eq: DSSF}
    \mathcal{S}^{ab}(\mathbf{q}, \omega)=&\frac{1}{N_{\text{d.u.c. }}} \sum_{\mathbf{R}_i, \mathbf{R}_j^{\prime}} P^{ab}_{ij}(\mathbf{q}) e^{i \mathbf{q} \cdot\left(\mathbf{R}_i-\mathbf{R}_j^{\prime}\right)} \nonumber \\
    &\hspace{1cm} \times \int d t e^{i \omega t}\left\langle\mathrm{~S}_{\mathbf{R}_i}^a(t) \mathrm{S}_{\mathbf{R}_j^{\prime}}^b(0)\right\rangle,
\end{align}
which is directly relevant for momentum-resolved inelastic neutron scattering experiments. With unpolarized neutrons, the prefactor denotes the transverse projector  
\begin{align}\label{eq: transverse projector DSSF}
    P^{ab}_{ij}(\mathbf{q}) = \left[\bm{\hat{e}}_{i}^a \cdot \bm{\hat{e}}_{j}^{b} - \frac{\left(\bm{\hat{e}}_{i}^{a} \cdot \mathbf{q}\right)\left(\bm{\hat{e}}_{j}^{b} \cdot \mathbf{q}\right)}{|\mathbf{q}|^2}\right],
\end{align}
where $\bm{\hat{e}}_{i}^a$ are basis vectors (i.e., $a=x,y,z$) for the local spin coordinates at the $i=0,1,2,3$ pyrochlore sublattice (see Appendix~\ref{Appendix Sec:Conventions}). 

\begin{figure}
\centering
\includegraphics[width=1.0\linewidth]{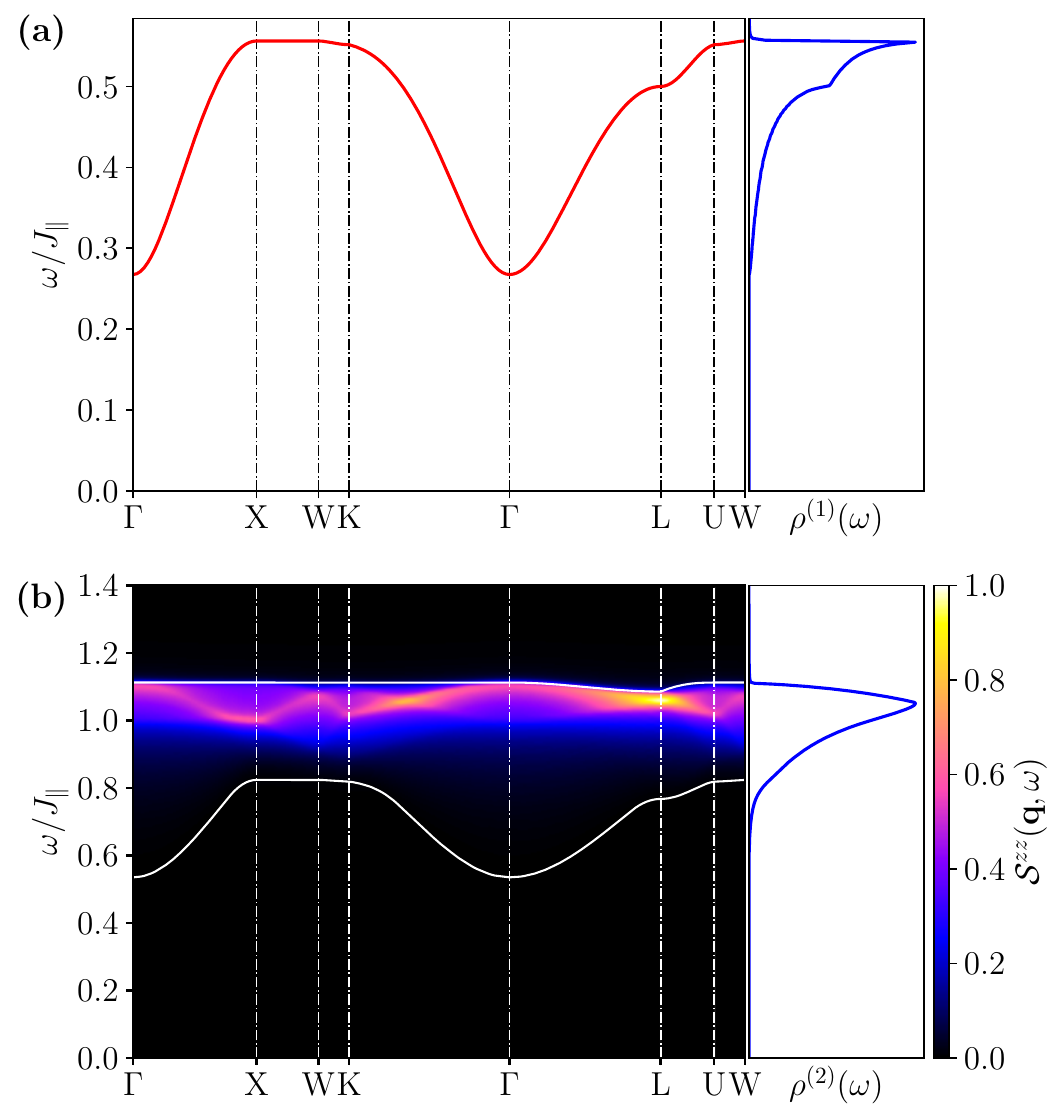}
\caption{(a) Spinon dispersion and corresponding single spinon density of states $\rho^{(1)}(\omega)$, and (b) dynamical spin structure factor $S^{zz}(\mathbf{q},\omega)$ (assuming $J_{\parallel}=J_{xx}$ or $J_{\parallel}=J_{yy}$) and two-spinon density of states $\rho^{(2)}(\omega)$ for 0-flux QSI with $J_{\pm}/J_{\parallel}=0.06$. White lines represent the lower and upper edges of the two-spinon continuum. \label{fig:dssf_0}}
\end{figure}

In the following, we will report $\mathcal{S}^{zz}(\mathbf{q},\omega)$ since these are the dynamical correlations that are directly relevant for inelastic neutron scattering on dipolar-octupolar candidates. In dipolar-octupolar systems, two of the pseudospin components ($\mathrm{S}^{x}$,$\mathrm{S}^{z}$) transform as dipoles and one ($\mathrm{S}^{y}$) as an octupole. This implies that the most general symmetry allowed Hamiltonian with nearest-neighbor coupling can be written as an XYZ model $\mathcal{H}=\sum_{\left\langle\mathbf{R}_i \mathbf{R}_j^{\prime}\right\rangle}\left[J_{x x} \mathrm{S}_{\mathbf{R}_i}^x \mathrm{S}_{\mathbf{R}_j^{\prime}}^x+J_{y y} \mathrm{S}_{\mathbf{R}_i}^y \mathrm{S}_{\mathbf{R}_j^{\prime}}^y+J_{z z} \mathrm{S}_{\mathbf{R}_i}^z \mathrm{S}_{\mathbf{R}_j^{\prime}}^z\right]$ and that the magnetic field linearly couples only to $\mathrm{S}^{x}$ and $\mathrm{S}^{z}$. However, $g_{xx}\approx 0$ due to the underlying octupolar magnetic charge density of $\mathrm{S}^{x}$~\cite{huang2014quantum, sibille2020quantum}. Neutron scattering at small momentum transfer is then only sensitive to $\expval{\mathrm{S}^z \mathrm{S}^z}$ correlations. In the case where the dominant interaction is between $\mathrm{S}^z$ components (i.e., $J_{\parallel}=J^{zz}$), this implies that neutrons only see the emergent photon $\expval{\mathrm{S}^z \mathrm{S}^z}\sim \expval{E E}$ since $\mathrm{S}^z\sim E$ from the above mapping to emergent quantum electrodynamics. In contrast, if the dominant coupling in the XYZ model is between the $\mathrm{S}^x$ or $\mathrm{S}^z$ components (i.e., $J_{\parallel}=J_{xx}$ or $J_{\parallel}=J_{zz}$) neutrons scattering would probe the two-spinon continuum $\expval{\mathrm{S}^z \mathrm{S}^z}\sim \expval{\Phi^\dagger \Phi \Phi^\dagger \Phi}$. In the former case, the spin structure factor can be computed using Gaussian quantum electrodynamics, as explained in Ref.~\cite{benton2012seeing}. In the latter, GMFT (using the large-$N$ or SCEBR) provides an analytical tool to make predictions about measurements. Experiments on dipolar-octupolar compounds Ce$_2$Zr$_2$O$_7$~\cite{smith2022case, Beare2923MuSr, smith2023quantum, bhardwaj2022sleuthing}, Ce$_2$Sn$_2$O$_7$~\cite{sibille2015candidate, sibille2020quantum, poree2023fractional}, and Ce$_2$Hf$_2$O$_7$~\cite{poree2023dipolar, Poree2022Crystal} indicate that, in all of them, the dominant coupling is either $J_{xx}$ or $J_{yy}$. We will similarly assume that $J_{\parallel}=J_{xx}$ or $J_{\parallel}=J_{yy}$ (i.e., $\expval{\mathrm{S}^z \mathrm{S}^z}$ is independent of this choice because of the $U(1)$ symmetry of the XXZ model). The $\expval{\mathrm{S}^z \mathrm{S}^z}$ correlations are then determined by the spinons and directly relevant to experiments on dipole-octupole systems. Detailed derivation and analytical expressions for the dynamical spin structure factor at finite temperature are given in Appendix~\ref{Appendix Sec: spin structure factor}.

\begin{figure}
\centering
\includegraphics[width=1.0\linewidth]{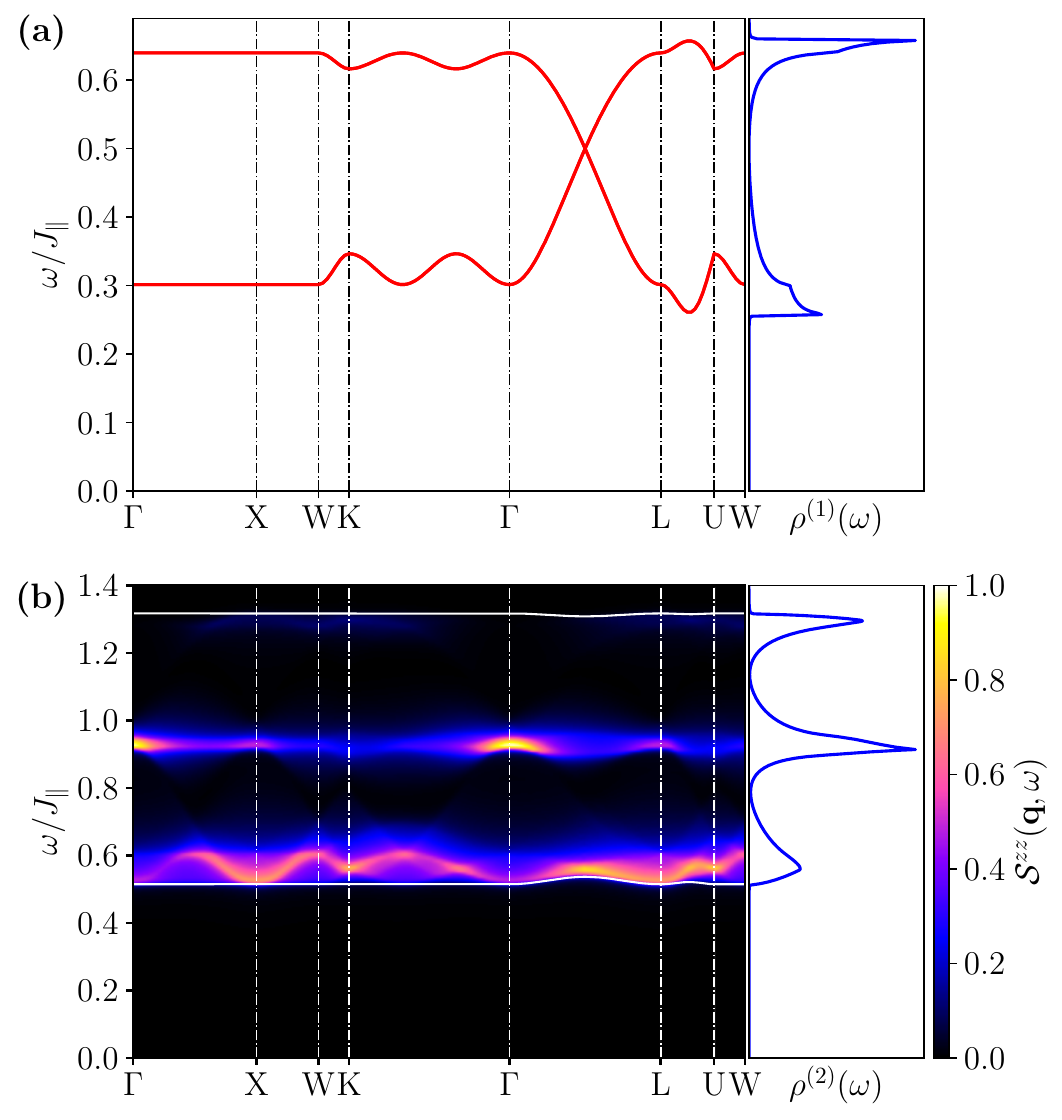}
\caption{(a) Spinon dispersion and corresponding single spinon density of states $\rho^{(1)}(\omega)$, and (b) dynamical spin structure factor $S^{zz}(\mathbf{q},\omega)$ (assuming $J_{\parallel}=J_{xx}$ or $J_{\parallel}=J_{yy}$) and two-spinon density of states $\rho^{(2)}(\omega)$ for $\pi$-flux QSI with $J_{\pm}/J_{\parallel}=-0.2$. White lines represent the lower and upper edges of the two-spinon continuum. \label{fig:dssf_pi}}
\end{figure}

Fig.~\ref{fig:dssf_0} presents the spinon dispersion and $\mathcal{S}^{zz}(\mathbf{q},\omega)$ for 0-QSI. Alongside these results, the single spinon density of states (DOS)
\begin{align}
    \rho^{(1)}(\omega) = \sum_{\mathbf{k},m}\delta(\omega-\varepsilon_{m}(\mathbf{k}))/N_{\text{u.c.}}
\end{align}
and the two-spinon density of states 
\begin{align}
    \rho^{(2)}(\omega) &= \sum_{\mathbf{k},\mathbf{q},m,n}\delta(\omega - \varepsilon_{m}(\mathbf{k}) - \varepsilon_{n}(\mathbf{q}))/N_{\text{u.c.}}^2 \label{eq: two-spinon dos} \\
    &=\int d\Omega \rho^{(1)}(\Omega)\rho^{(1)}(\omega-\Omega) \label{eq: two-spinon dos convolution}
\end{align}
are also shown. In the above equations, $m,n$ are band indices, $N_{\text{u.c.}}$ is the number of GMFT unit cells, and the sum is over the associated reduced first Brillouin zone. As already mentioned, the spinon dispersion has a single band with a minimum at the zone center. The associated dynamical spin structure factor produces a broad signal with most of the spectral weight close to the upper edge of the two-spinon continuum. This observation can be understood by noticing that the two-spinon DOS is negligible close to the lower edge and reaches its maximum around the upper edge of the continuum. These results are similar to the ones obtained in the large-$N$~\cite{desrochers2023spectroscopic} approach and also consistent with the QMC results of Ref.~\cite{huang2018dynamics}.

For $\pi$-QSI, the spinon dispersion and dynamical correlations are reported in Fig.~\ref{fig:dssf_pi}. The two primarily flat bands can be clearly observed. Associated with this dispersion, the single spinon DOS has two peaks centered around the flat bands and vanishes at intermediate energy, where the two bands meet at Dirac points. The dynamical spin structure factor has three peaks as a function of energy for a given momentum transfer (the third peak close to the upper edge of the two-spinon continuum is faint). These peaks are a smoking-gun signature unique to $\pi$-QSI as highlighted in Ref.~\cite{desrochers2023spectroscopic}. The three-peak structure is a direct consequence of the two flat spinon bands. To understand this observation, one should recall that the dynamical spin structure factor is weighted by the two-spinon DOS (see Appendix~\ref{Appendix Sec: spin structure factor}). Using the definition of the two-spinon DOS~\eqref{eq: two-spinon dos}, it can be seen that there are three main contributions $\rho^{(2)}(\omega)\sim \sum_{\mathbf{k},\mathbf{q}}\delta(\omega-\varepsilon_{-}(\mathbf{k})-\varepsilon_{-}\mathbf{q}) + \delta(\omega-\varepsilon_{-}(\mathbf{k})-\varepsilon_{+}\mathbf{q}) - \delta(\omega-\varepsilon_{+}(\mathbf{k})-\varepsilon_{+}\mathbf{q})$. The first contribution with two spinons in the lower band yields the lowest-energy peak, the second one with two spinons in different bands corresponds to the intermediate peaks, and the third one is due to two spinons in the second spinon band. 
Recent experiments on candidate material Ce$_2$Sn$_2$O$_7$ have reported hints of this three-peak structure using backscattering neutron spectroscopy on powder samples~\cite{poree2023fractional}. 

\begin{figure}
\centering
\includegraphics[width=1.0\linewidth]{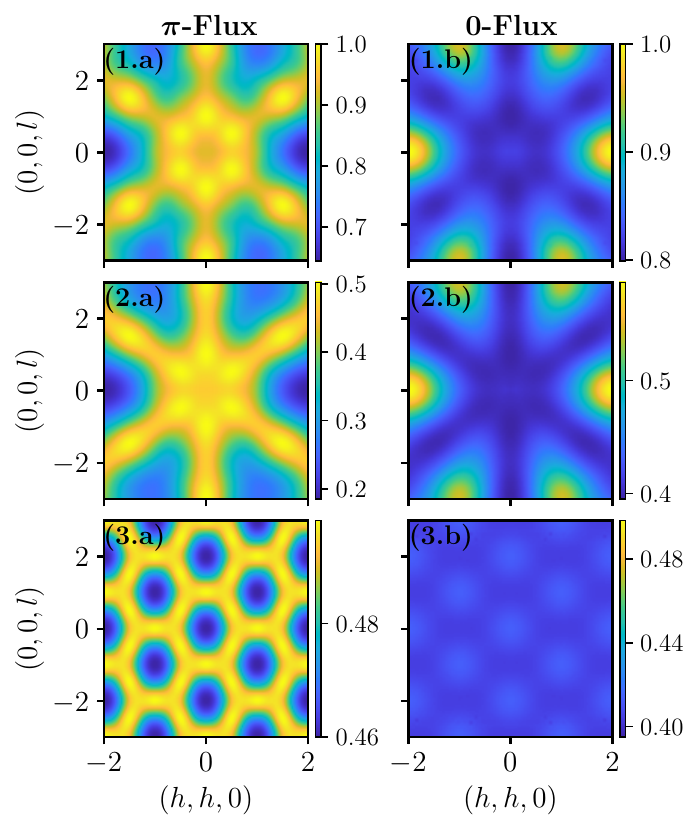}
\caption{(1) Equal-time structure factor $\mathcal{S}^{zz}(\mathbf{q})$ in the $(h,h,l)$ plane and its contribution to the (2) spin-flip and (3) non-spin flip channel with neutrons polarized perpendicular to the scattering plane for (a) $\pi$-flux QSI with $J_{\pm}/J_{\parallel}=-0.2$ and (b) 0-flux QSI with $J_{\pm}/J_{\parallel}=0.06$. \label{fig:sssf}}
\end{figure}

Fig.~\ref{fig:sssf} presents the equal-time (energy-integrated) structure factor $\mathcal{S}^{zz}(\mathbf{q})=\int d\omega \mathcal{S}^{zz}(\mathbf{q},\omega)$ in the $(h,h,l)$ plane for 0- and $\pi$-QSI. It also reports the contribution to the equal-time structure factor for the spin-flip $\mathcal{S}^{zz}_{\text{SF}}(\mathbf{q})$ and non-spin-flip $\mathcal{S}^{zz}_{\text{NSF}}(\mathbf{q})$ channels with neutron polarized perpendicular to the scattering plane. These quantities are relevant to energy-integrated inelastic neutron scattering with polarized neutrons. $\mathcal{S}^{zz}_{\text{SF}}(\mathbf{q})$ and $\mathcal{S}^{zz}_{\text{NSF}}(\mathbf{q})$ are calculated using Eq.~\eqref{eq: DSSF} using the prefactors $P^{zz}_{ij,\text{SF}}(\mathbf{q}) = \left(\bm{\hat{e}}_{i}^z \cdot \bm{\hat{z}}^{\text{sc}}\right) \left(\bm{\hat{e}}_{j}^z \cdot \bm{\hat{z}}^{\text{sc}}\right)$ and $P^{zz}_{ij,\text{NSF}}(\mathbf{q}) = \left(\bm{\hat{e}}_{i}^z \cdot \frac{\mathbf{q}\times \bm{\hat{z}}^{\text{sc}}}{|\mathbf{q}\times \bm{\hat{z}}^{\text{sc}}|} \right) \left(\bm{\hat{e}}_{j}^z \cdot \frac{\mathbf{q}\times \bm{\hat{z}}^{\text{sc}}}{|\mathbf{q}\times \bm{\hat{z}}^{\text{sc}}|} \right)$, respectively, where $\bm{\hat{z}}^{\text{sc}}$ is a unit vector perpendicular to the scattering plane.

When examining the figure, it should be noticed that the equal-time structure factor for 0- and $\pi$-flux show inverse patterns in intensity. This offers another way to distinguish 0- and $\pi$-QSI experimentally. In particular, $\pi$-flux (0-flux) QSI displays a high-intensity (low-intensity) snowflake-like pattern (or so-called rod motifs~\cite{castelnovo2019rod}) in $\mathcal{S}^{zz}(\mathbf{q})$ and $\mathcal{S}^{zz}_{\text{SF}}(\mathbf{q})$. The non-spin-flip channel of the $\pi$-flux (0-flux) state shows strong scattering along the zone boundaries (zone center). This inverted intensity can be understood by noticing that at the Ising point, the dominant spin component correlations $\expval{S^{\parallel}S^{\parallel}}$ show strong classical spin ice correlation (e.g., pinch points~\cite{castelnovo2012spin, fennell2009magnetic, bramwell2001spin}), whereas the two transverse correlators ($\expval{S^{z} S^{z}}$ and $\expval{S^{y} S^{y}}$ if $J_{\parallel}=J_{xx}$ or $\expval{S^{z} S^{z}}$ and $\expval{S^{x} S^{x}}$ if $J_{\parallel}=J_{yy}$) are completely flat and featureless in momentum space. When the transverse coupling $J_{\pm}$ becomes non-zero, the dominant correlations remain mostly unaffected, but the transverse ones develop ferromagnetic (antiferromagnetic) correlations for $J_{\pm}<0$ ($J_{\pm}>0$). Since $\expval{S^{z} S^{z}}$ is mainly determined by $J_{\pm}$, it is natural to understand why it essentially shows the opposite behavior when $J_{\pm}$ changes sign. The SCEBR predictions for the equal-time structure factor of $\pi$-QSI should also be compared with energy-integrated inelastic neutron scattering measurements on candidate material Ce$_2$Zr$_2$O$_7$~\cite{gaudet2019quantum, smith2022case}. The two are strikingly similar.


\section{\label{sec: Finite temperature results} Finite temperature results}

\begin{figure}
\centering
\includegraphics[width=1.0\linewidth]{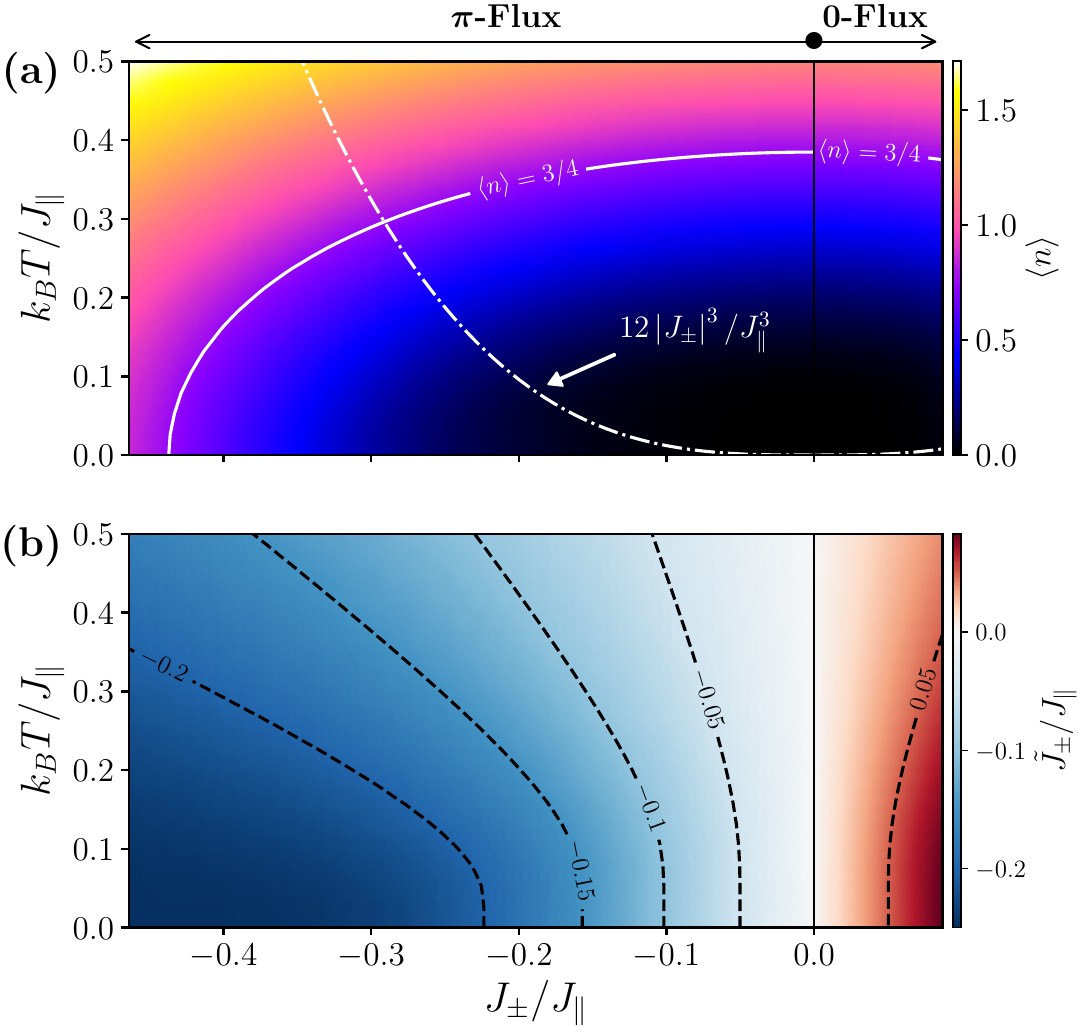}
\caption{(a) Average tetrahedron occupation as a function of temperature and transverse coupling. Also shown is the approximate scaling of the vison gap $\Delta_{\text{vison}}\sim 12|J_{\pm}|^3/J_{\parallel}^2$ and a contour line for the tetrahedron occupation in the $T\to\infty$ limit $\expval{n}=3/4$. (b) The associated renormalized transverse coupling $\tilde{J}_{\pm}=J_{\pm}/(1+\expval{n})$. \label{fig:Finite temperature density}}
\end{figure}

\subsection{Thermodynamics and crossovers}

Another advantage of the SCEBR formulation is that it can be straightforwardly generalized to study QSI at finite temperatures.
Instead of using a ground state expectation value when computing $\expval{n}$ in the self-consistency condition~\eqref{eq: renormalized transverse projector}, one simply has to use a thermal average. As an example of results that can be obtained with the formulation, Fig.~\ref{fig:Finite temperature density}(a) presents the average tetrahedron occupation as a function of temperature and transverse coupling for both 0- and $\pi$-QSI. As expected, we see that the tetrahedron occupation slowly rises as a function of temperature up to a point where it even goes beyond the $T\to\infty$ limit $\expval{n}=3/4$. The fact that the occupation goes beyond this value is an artifact of the theory because the exclusiveness~\eqref{eq:exclusiveness constraint} and maximum occupation constraints~\eqref{eq:occupation constraint} have been dropped. The contour line that indicates when the system reaches $\expval{n}=3/4$ is an approximate indicator of the temperature where the system crosses over from a constrained (cooperative) to a trivial high-temperature paramagnet. This estimate for the crossover position is of the same order of magnitude as in QMC, where the crossover region starts at about $k_BT/J_{\parallel}=1$ and ends around $k_BT/J_{\parallel}=0.1$ for 0-QSI~\cite{kato2015numerical, huang2018dynamics}. The $\expval{n}=3/4$ line also provides an estimate for the range of validity of the SCEBR since the theory should only be valid at temperatures below this limit. Fig.~\ref{fig:Finite temperature density}(a) also shows the estimated scaling of the vison gap (i.e., magnetic monopole of the emergent $U(1)$ gauge theory) as a function of transverse coupling $\Delta_{\text{vison}}\sim 12 |J_{\pm}|^3/J_{\parallel}^2$~\cite{hermele2004pyrochlore, Motrunich2005Origin, bergman2006ordering, Szabo2019Seeing, Kwasigroch2020Vison, sanders2023vison}. Above this energy scale, thermally excited gauge fluxes should become increasingly important. Such a graphical representation of the different energy scales at play highlights crucial differences between 0- and $\pi$-QSI at finite temperatures. 

The finite temperature behavior of $0$-flux QSI is already well-established~\cite{banerjee2008unusual, kato2015numerical, huang2018dynamics, huang2020extended}. As the system is cooled down, it undergoes two successive crossovers at temperatures that we will denote by $T_1$ and $T_2$. These are characterized by (non-singular) peaks in the heat capacity. The first one occurs around $k_B T_1=\mathcal{O}(J_{\parallel})$, and marks a transition from a trivial paramagnet to classical spin ice (i.e., a constrained or cooperative paramagnet). At this crossover, the entropy per site goes from the high-temperature limit of $S=k_B\ln(2)$ to a plateau around Pauling's entropy $S = k_B \ln (3 / 2) / 2$. This signals that, after the crossover, the system is energetically constrained to the spin ice manifold and is described by thermal fluctuations within these 2-in-2-out states. At the second crossover temperature controlled by the vison gap $k_B T_2=\mathcal{O}(J_{\pm}^3/J_{\parallel}^2)$, the entropy is quenched $S\to 0$ as the system transitions from a classical to a quantum spin liquid with genuine deconfined fractional excitations and emergent gauge fields. After this temperature, the fluxes are frozen, and the dynamic of the system is controlled by quantum coherent effects rather than thermal fluctuations. Because the 0-flux phase is only stable for small values of the transverse coupling, the vison gap is always much smaller than the leading coupling. Consequently, the first and second crossover are well separated (i.e., $k_B T_1\gg k_B T_2$). This is clearly seen in Fig.~\ref{fig:Finite temperature density}(a) where the $\expval{n}=3/4$ contour line is always at a much higher temperature than the approximate vison gap for $J_{\pm}>0$.

This should be contrasted with $\pi$-QSI. Due to its enhanced stability to the transverse coupling, there is a significant parameter regime where the vison gap and leading coupling $J_{\parallel}$ are of the same order of magnitude. This has important consequences. First, it implies that one can reach the QSI regime ($k_B T < k_BT_1$ and $k_B T < k_BT_2$) at much higher temperatures relative to the microscopic exchange scale $J_{\parallel}$. This makes the prospect of experimentally accessing the QSI regime with currently available methods much more promising in the $\pi$-flux than the 0-flux phase. Second, the similar energy scale of the vison and spinon gap implies that one should expect a single crossover directly from the trivial paramagnet to the QSI. In thermodynamic measurements, one should then expect a single broad non-singular peak in the heat capacity where the entropy per site goes from $S=\ln(2)$ to zero without any intermediate plateau. Of course, the vison gap should be significantly renormalized compared to the naive scaling illustrated in Fig.~\ref{fig:Finite temperature density} for large values of $|J_{\pm}|/J_{\parallel}$. Nevertheless, we expect the above arguments to remain valid.


\begin{figure}
\centering
\includegraphics[width=1.0\linewidth]{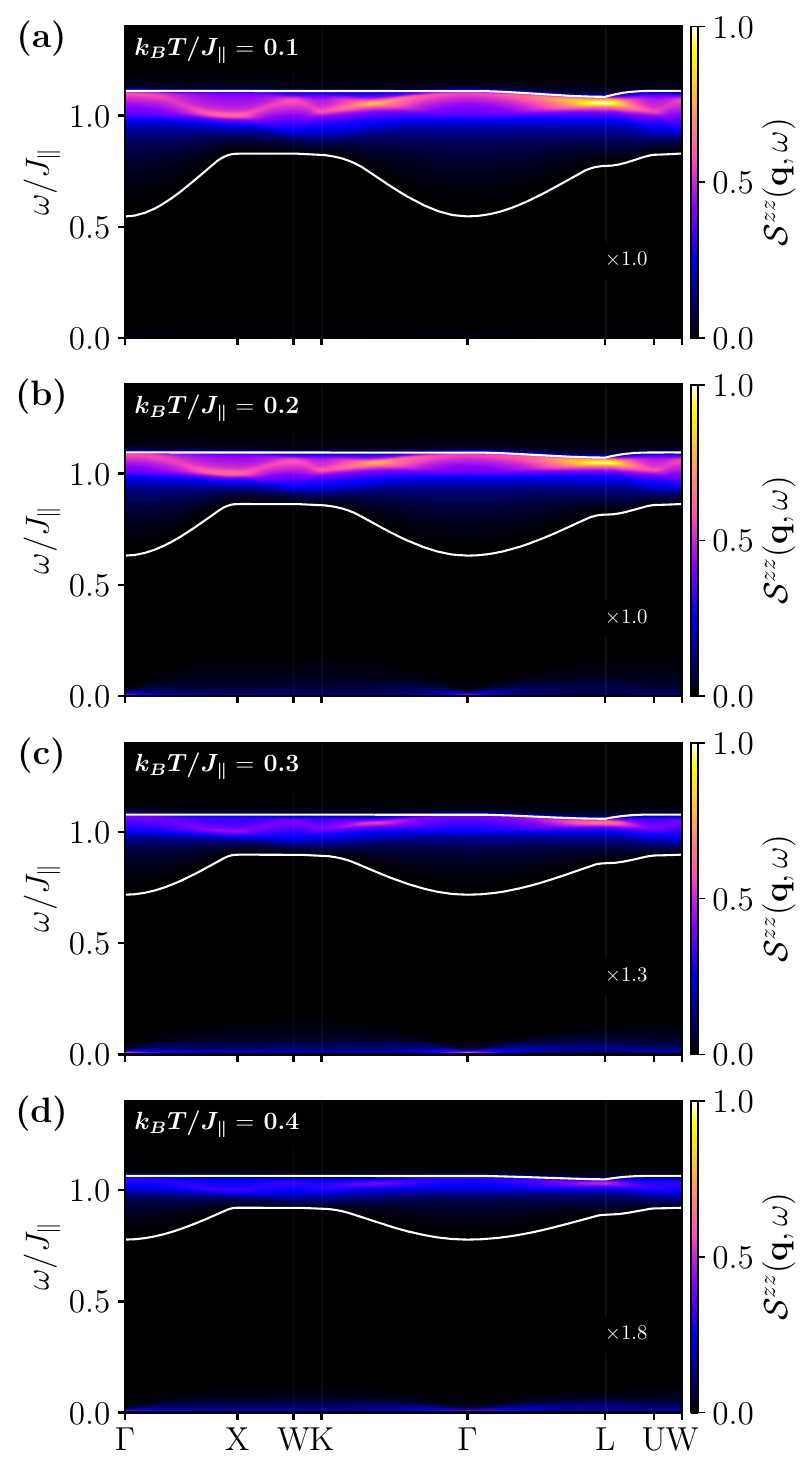}
\caption{Finite temperature dynamical spin structure factor $S^{zz}(\mathbf{q},\omega)$ (assuming $J_{\parallel}=J_{xx}$ or $J_{\parallel}=J_{yy}$) for 0-QSI with $J_{\pm}/J_{\parallel}=0.06$ at (a) $k_B T/J_{\parallel}=0.1$, (b) $k_B T/J_{\parallel}=0.2$, (c) $k_B T/J_{\parallel}=0.3$, and (d) $k_B T/J_{\parallel}=0.4$. The multiplicative factor is the scale of the colorbar with respect to results at $k_B T/J_{\parallel}=0.1$. \label{fig:Finite temperature dssf 0}}
\end{figure}

\begin{figure}
\centering
\includegraphics[width=1.0\linewidth]{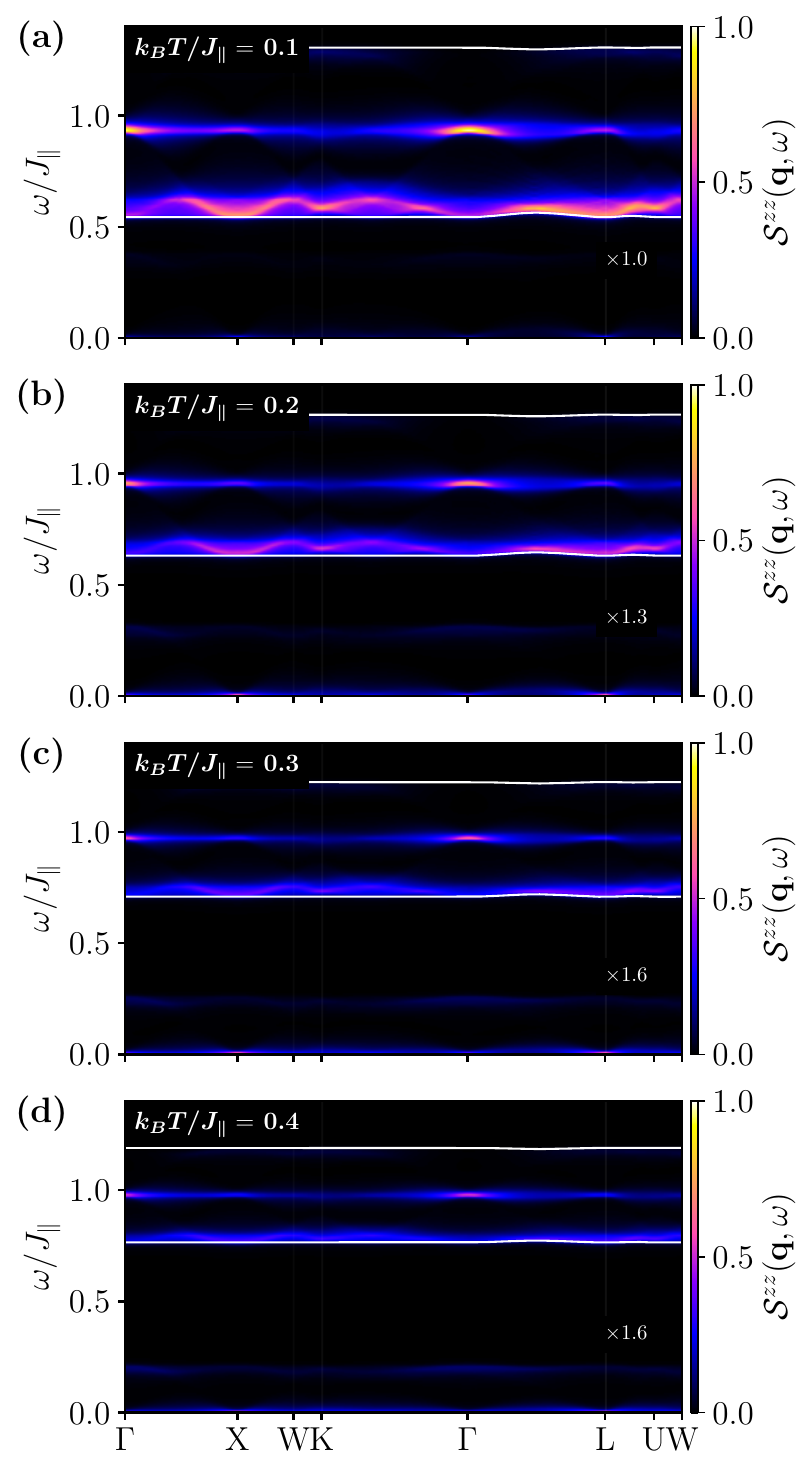}
\caption{Finite temperature dynamical spin structure factor $S^{zz}(\mathbf{q},\omega)$ (assuming $J_{\parallel}=J_{xx}$ or $J_{\parallel}=J_{yy}$) for $\pi$-QSI with $J_{\pm}/J_{\parallel}=-0.2$ at (a) $k_B T/J_{\parallel}=0.1$, (b) $k_B T/J_{\parallel}=0.2$, (c) $k_B T/J_{\parallel}=0.3$, and (d) $k_B T/J_{\parallel}=0.4$. The multiplicative factor is the scale of the colorbar with respect to results at $k_B T/J_{\parallel}=0.1$. \label{fig:Finite temperature dssf 1}}
\end{figure}

\subsection{Finite temperature spinon dynamics}

Fig.~\ref{fig:Finite temperature density}(b) presents the associated renormalized transverse coupling as a function of temperature. Because of the rise in the average tetrahedron occupation, the renormalized transverse coupling slowly decreases as the temperature is increased for a given initial value of $J_{\pm}/J_{\parallel}$. Transverse coupling mediates spinon hopping in the XXZ model. The above thus describes a thermal inhibition of spinon hopping processes. Indeed, a high density of tetrahedron occupancy should disfavor spinon hopping since it reduces the number of hopping processes possible, and double tetrahedron occupancy is energetically disfavored. The SCEBR thus captures this effect in an average way through a self-consistent reduction of the transverse coupling. 

To see the potential physical consequences of this thermally induced spinon hopping suppression, we present the dynamical spin structure factor $\mathcal{S}^{zz}(\mathbf{q},\omega)$ at finite temperatures up to $k_B T/J_{\parallel} = 0.4$ in Figs.~\ref{fig:Finite temperature dssf 0} and~\ref{fig:Finite temperature dssf 1} for 0- and $\pi$-QSI, respectively. As already mentioned, the SCEBR should not remain valid up to these very high temperatures. Nevertheless, we push through in order to see what potential effects the theoretical formulation captures/misses at these finite temperatures. The first effect which is immediately noticeable upon inspection of these figures is the reduction of the energy range where $\mathcal{S}^{zz}(\mathbf{q},\omega)$ is non-zero (i.e., the width of the two-spinon continuum) with increasing temperature. Crucially, this thermal suppression of the spinon bandwidth has been seen in QMC simulation on 0-QSI. Ref.~\cite{huang2018dynamics} reported that the bandwidth of the transverse dynamical spin structure factor decreases when temperature increases. In this QMC study, the continuum also evolves from having clear local maxima at specific high-symmetry points to mostly featureless and momentum-independent. The SCEBR thus offers an analytical approach that captures some of these effects and provides a clear underlying mechanism.

Nonetheless, we emphasize that important physical effects are clearly missed due to the approximations that go into the SCEBR of GMFT. Most notably, The approach does not consider thermally excited fluxes that should undoubtedly play an important role at high temperatures. In a sense, GMFT only considers the average (0 or $\pi$) flux background while neglecting any fluctuation around it. It is known that the presence of random thermal fluxes can also lead to the localization of fractional excitation in spin liquids~\cite{Hart2020Coherent, hart2021correlation, Kim2022Anderson, Udagawa2019Spectrum}. Thermal fluxes thus also provide an additional explanation for the suppression of the continuum bandwidth reported in QMC but are not considered by our current description.

Another notable effect seen in Figs.~\ref{fig:Finite temperature dssf 0} and~\ref{fig:Finite temperature dssf 1}, is the emergence of a quasielastic signal close to $\omega=0$. The physical origin of this signal is most explicit by remarking that the dynamical spin structure factor is of the generic form (see Appendix~\ref{Appendix Sec: spin structure factor}) $S^{zz}(\mathbf{q},\omega)\sim \sum_{\mathbf{k},m,n}(1+n_B(\varepsilon_{m}(\mathbf{k})))(1+n_B(\varepsilon_{m}(\mathbf{k}-\mathbf{q})))\delta(\omega - \varepsilon_m(\mathbf{k}) - \varepsilon_n(\mathbf{k}-\mathbf{q})) + n_B(\varepsilon_{m}(\mathbf{k}))(1+n_B(\varepsilon_{m}(\mathbf{k}-\mathbf{q})))\delta(\omega + \varepsilon_m(\mathbf{k}) - \varepsilon_n(\mathbf{k}-\mathbf{q})) + n_B(\varepsilon_{m}(\mathbf{k})) n_B(\varepsilon_{m}(\mathbf{k}-\mathbf{q}))\delta(\omega + \varepsilon_m(\mathbf{k}) + \varepsilon_n(\mathbf{k}-\mathbf{q}))$, where $n_{B}$ is the Bose-Einstein distribution. At zero temperature, when the ground state is a spinon vacuum, the only allowed process in inelastic neutron scattering is the scattering of the neutron by the creation of a spinon pair from the vacuum (i.e., $\delta(\omega - \varepsilon_m(\mathbf{k}) - \varepsilon_n(\mathbf{k}-\mathbf{q}))$). Accordingly, $\mathcal{S}^{zz}(\mathbf{q},\omega)$ is only non-zero within the two-spinon continuum. At finite temperature, there is a finite density of thermally excited spinons. A neutron can then be scattered by deexciting a thermal spinon while exciting another one (i.e., $\delta(\omega + \varepsilon_m(\mathbf{k}) - \varepsilon_n(\mathbf{k}-\mathbf{q}))$) for an approximately vanishing net energy transfer. It is exactly this kind of process that leads to the quasielastic contribution seen in the dynamical spin structure factor.

\section{\label{sec: Discussion} Discussion}

In this work, we have introduced the self-consistent exclusive boson representation (SCEBR). The SCEBR is an analytical description of spinon excitations in QSI based on the GMFT construction that, rather than using a large-$N$ approximation, represents the $O(2)$ quantum rotor by a set of two mutually exclusive bosons as in Ref.~\cite{gingras2014quantum}. We have shown that the low-density approximation initially introduced fails for the $\pi$-flux phase beyond the perturbative Ising limit. To mediate these issues and extend the range of validity of the exclusive boson construction, the effect of a  finite density of emergent charges can be treated in an average way. The essential element of the SCEBR is a self-consistent reduction of spinon hopping caused by the presence of neighboring emergent charges.

We have then explored the formalism and its physical predictions in detail. We have used analytical expressions for the spinon dispersion and Bogoliubov transformation to compute the dynamical and equal-time spin structure factor of 0- and $\pi$-QSI. This was used to show that the spinon contribution to the DSSF has a broad continuum with a single peak as a function for energy for the 0-flux state in contrast to three sharp peaks for $\pi$-flux --- in agreement with previous large-$N$ treatment~\cite{desrochers2023spectroscopic}. The equal-time structure factors relevant for dipolar-octupolar systems $\mathcal{S}^{zz}(\mathbf{q})$ display snowflake patterns in the $(h,h,l)$ that are opposite in intensity for 0- and $\pi$-QSI, consistent with ED~\cite{Hosoi2022Uncovering} and pseudofermion functional renormalization group~\cite{chern2023pseudofermion}. We then moved to discuss finite temperature results. Our approach highlighted apparent differences in thermodynamic expectation for the $\pi$- and 0-flux cases. There should be a single crossover from a trivial paramagnet to the QSL for $\pi$-QSI (beyond the perturbative Ising limit) in contrast to two crossovers with an intermediate classical spin ice regime for 0-QSI. Examination of spinon dynamics at finite temperature further highlighted that the SCEBR predicts a reduction of the two-spinon continuum width with temperature due to hopping suppression by thermally excited spinons. This effect was also previously noted in QMC simulation on the 0-QSI.

The SCEBR thus offers an analytical approach that has been benchmarked and can be readily used to study material QSI candidates. It should further be noted that the SCEBR has several advantages over the usual large-$N$ approach to GMFT. It is more analytically tractable, and its expected range of validity is more evident due to its transparent physical motivation. Separating the positive and negative charges makes the physical interpretation and calculations of different processes more explicit. It also yields more sound results when studying finite temperature phenomena~\cite{savary2013spin}. A final advantage already emphasized in Ref.~\cite{hao2014bosonic} is that the method is more straightforwardly amenable to improvement via standard diagrammatic many-body treatments than the large-$N$ approach. 

The SCEBR opens the door to many theoretical investigations of interest. The model could be used to study the XYZ model or any other more general microscopic model of interest rather than the simple XXZ considered in this article. It could be interesting to see how the phase diagram predicted by the SCEBR differs from the ones obtained for the symmetry-allowed model relevant to the dipolar-octupolar~\cite{desrochers2023spectroscopic}, the effective spin-1/2~\cite{savary2012coulombic, savary2013spin}, and the Non-Kramers cases~\cite{lee2012generic}. Similarly, one could look at how predictions for the equal-time and dynamical correlations differ from results obtained using the large-$N$ and other methods for these more general models. 

On top of being used in more general and complex situations, the formalism could also be extended in many significant ways. One could explicitly consider spinon-gauge interactions by expanding gauge fluctuations around the saddle point (i.e., $A\to \overline {A} + \tilde{A}$) similarly to Refs.~\cite{hao2014bosonic, Fu2017Fingerprints, Seth2022Probing}. When solving the self-consistency condition in the SCEBR, one would have to use the full interacting Green's function rather than the non-interacting one. It would be interesting to see how predictions for the phase diagram and dynamical properties differ in the interacting and non-interacting cases. As mentioned in the main text, an important aspect neglected in the current formulation is the presence of thermally excited fluxes that should play an important role at finite temperatures. It would be highly desirable to construct a model that could incorporate these to have more accurate models of QSI at finite temperatures. This is especially true for $\pi$-QSI, where physical insight is much harder to develop due to the sign problem in QMC. The SCEBR could potentially be a starting point for such a non-trivial generalization. More broadly, we hope this alternative approach can help guide the search for QSI materials by offering a tool to help interpret measurements and potentially highlight novel experimental signatures. 

\begin{acknowledgments}
We acknowledge support from the Natural Sciences and Engineering Research Council of Canada (NSERC) and the Centre of Quantum Materials at the University of Toronto. Computations were performed on the Niagara cluster, which SciNet hosts in partnership with the Digital Research Alliance of Canada. FD is supported by the Vanier Canada Graduate Scholarship. 
\end{acknowledgments}

\appendix

\section{\label{Appendix Sec:Conventions} Local coordinates}

Spins on the four different pyrochlore sublattices are defined in a local frame. The basis vectors of these sublattice-dependant coordinates systems are given in table~\ref{tab: Local basis}.

\begin{table}[!ht]
\caption{\label{tab: Local basis}%
Local sublattice basis vectors
}
\begin{ruledtabular}
\begin{tabular}{ccccc}
$i$ & 0 & 1  & 2  & 3 \\
\hline
$\bm{\hat{e}}_{i}^z$ & $\frac{1}{\sqrt{3}}\left(1,1,1\right)$ & $\frac{-1}{\sqrt{3}}\left(-1,1,1\right)$  & $\frac{-1}{\sqrt{3}}\left(1,-1,1\right)$  & $\frac{-1}{\sqrt{3}}\left(1,1,-1\right)$   \\[2mm]
$\bm{\hat{e}}_{i}^y$ & $\frac{1}{\sqrt{2}}\left(0,-1,1\right)$  & $\frac{1}{\sqrt{2}}\left(0,1,-1\right)$  & $\frac{-1}{\sqrt{2}}\left(0,1,1\right)$ & $\frac{1}{\sqrt{2}}\left(0,1,1\right)$  \\[2mm]
$\bm{\hat{e}}_{i}^x$ & $\frac{1}{\sqrt{6}}\left(-2,1,1\right)$ & $\frac{-1}{\sqrt{6}}\left(2,1,1\right)$  & $\frac{1}{\sqrt{6}}\left(2,1,-1\right)$  & $\frac{1}{\sqrt{6}}\left(2,-1,1\right)$    \\
\end{tabular}
\end{ruledtabular}
\end{table}

\section{\label{Appendix Sec:Analytical diagonalization of the quadratic Hamiltonians} Analytical diagonalization of the quadratic Hamiltonians}

\subsection{Generalities}

By Fourier transforming the field operator for the $b$ and $d$ bosons
\begin{subequations}
\begin{align}
    b_{\mathbf{r}_\alpha} = \frac{1}{\sqrt{N_{\text {u.c. }}}} \sum_{\mathbf{k}} b_{\mathbf{k},  \mathbf{r}_s, \alpha} e^{i\mathbf{k} \cdot \mathbf{r}_\alpha} \\
    d_{\mathbf{r}_\alpha} = \frac{1}{\sqrt{ N_{\text {u.c. }}}} \sum_{\mathbf{k}} d_{\mathbf{k}, \mathbf{r}_s, \alpha} e^{i \mathbf{k} \cdot \mathbf{r}_\alpha },
\end{align}
\end{subequations}
where $N_{\text{u.c.}}$ is the number of unit cells for a given GMFT Ansatz, and the sum is over the associated reduced first Brillouin zone, the Hamiltonian can be rewritten as 
\begin{align}
    \mathcal{H} = \sum_{\mathbf{k}} \sum_{\alpha\in\{A,B\}}\vec{\Psi}^{\dagger}_{\mathbf{k}, \alpha} H_{\alpha}(\mathbf{k}) \vec{\Psi}_{\mathbf{k}, \alpha} ,
\end{align}
where we have introduced the notation
\begin{align}
    \vec{\Psi}^{\dagger}_{\mathbf{k},\alpha} =& \left( d^{\dagger}_{\mathbf{k},1,\alpha},..., d^{\dagger}_{\mathbf{k},N_{s},\alpha}, b^{\dagger}_{\mathbf{k},1,\alpha},..., b^{\dagger}_{\mathbf{k},N_{s},\alpha}, \right.\nonumber \\
    &\quad\left. d_{-\mathbf{k},1,\alpha},..., d_{-\mathbf{k},N_{s},\alpha}, b_{-\mathbf{k},1,\alpha},..., b_{-\mathbf{k},N_{s},\alpha}\right). 
\end{align}
$N_{s}$ is the number of diamond unit cells in the GMFT unit cells of a specific Ansatz. $N_{s}=1$ and $N_{s}=4$ for $0$- and $\pi$-flux, respectively. 

In all cases, provided the $H_{\alpha}$ matrix is positive definite, we can diagonalize it with a Bogoliubov transformation $P_{\alpha}(\mathbf{k})$ of the form
\begin{align}
    \vec{\Psi}_{\mathbf{k},\alpha} &= P_\alpha(\mathbf{k}) \vec{\Gamma}_{\mathbf{k},\alpha},
\end{align}
where the new eigenmodes 
\begin{align}
    \vec{\Gamma}^{\dagger}_{\mathbf{k},\alpha} =& \left( \gamma^{\dagger}_{\mathbf{k},1,\alpha}, ..., \gamma^{\dagger}_{\mathbf{k},2N_{s},\alpha}, \gamma_{-\mathbf{k},1,\alpha}, ..., \gamma_{-\mathbf{k},2N_s,\alpha} \right)
\end{align}
are defined such that
\begin{align}
    &P_\alpha^{\dagger}(\mathbf{k}) H_{\alpha}(\mathbf{k}) P_\alpha(\mathbf{k}) \nonumber \\
    &= \text{diag}\left( \varepsilon_{1,\alpha}(\mathbf{k}), ..., \varepsilon_{N_s,\alpha}(\mathbf{k}), \varepsilon_{1,\alpha}(\mathbf{k}), ..., \varepsilon_{N_s,\alpha}(\mathbf{k}) \right)\nonumber \\
    &= \text{diag}\left( \omega_{1,\alpha}(\mathbf{k}), ..., \omega_{2N_s,\alpha}(\mathbf{k}) \right). 
\end{align}
To preserve the canonical commutation relation, the Bogoliubov transformation has to satisfy the para-unitary condition~\cite{colpa1978diagonalization}
\begin{equation} \label{eq: paraunitary constraint}
    P^{\dagger}_{\alpha}(\mathbf{k}) J P_{\alpha}(\mathbf{k})=J, \quad J=\left(\sigma^z \otimes \mathds{1}_{2 N_{S L}, 2 N_{S L}}\right).
\end{equation}
One can directly find the spectrum by computing the eigenvalues of $JH_{\alpha}$. The eigenvalues of this matrix are $\left( \varepsilon_{1,\alpha}(\mathbf{k}), ..., \varepsilon_{N_s,\alpha}(\mathbf{k}), -\varepsilon_{1,\alpha}(\mathbf{k}), ..., -
\varepsilon_{N_s,\alpha}(\mathbf{k}) \right)$.

To make calculations and interpretation more explicit, we can consider the different blocks of the Bogoliubov transformation
\begin{align}
    P_{\alpha}(\mathbf{k}) &=
    \left(\begin{array}{@{}cc@{}}
    P_{\alpha}^{11}(\mathbf{k}) & P_{\alpha}^{12}(\mathbf{k})  \\[0.5mm] 
    P_{\alpha}^{21}(\mathbf{k}) & P_{\alpha}^{22}(\mathbf{k}) 
    \end{array}\right).
\end{align}
These different blocks satisfy
\begin{subequations}
    \begin{align}
        \left(P_\alpha^{11}(\mathbf{k})\right)^* &= P_\alpha^{22}(-\mathbf{k}) \\
        \left(P_\alpha^{12}(\mathbf{k})\right)^* &= P_\alpha^{21}(-\mathbf{k}),
    \end{align}
\end{subequations}
and can be used to conveniently rewrite the initial $b$ and $d$ bosons in terms of the $\gamma$ eigenmodes as 
\begin{subequations}
    \begin{align}
        b_{\mathbf{k},\mathbf{r}_s,\alpha} &= \sum_{\sigma=1}^{2N_s} \left( \left[P_{\alpha}^{11}(\mathbf{k})\right]_{(b,r_s),\sigma} \gamma_{\mathbf{k},\sigma,\alpha} \right.\nonumber \\
        &\hspace{1.5cm} \left. + \left[P_{\alpha}^{12}(\mathbf{k})\right]_{(b,r_s),\sigma} \gamma^{\dagger}_{-\mathbf{k},\sigma,\alpha}  \right)\\
        d_{\mathbf{k},\mathbf{r}_s,\alpha} &= \sum_{\sigma=1}^{2N_s} \left( \left[P_{\alpha}^{11}(\mathbf{k})\right]_{(d,r_s),\sigma} \gamma_{\mathbf{k},\sigma,\alpha} \right.\nonumber \\
        &\hspace{1.5cm} \left. + \left[P_{\alpha}^{12}(\mathbf{k})\right]_{(d,r_s),\sigma} \gamma^{\dagger}_{-\mathbf{k},\sigma,\alpha}  \right)\\
        b_{\mathbf{k},\mathbf{r}_s,\alpha}^{\dagger} &= \sum_{\sigma=1}^{2N_s} \left( \left[P_{\alpha}^{11}(\mathbf{k})\right]^*_{(b,r_s),\sigma} \gamma_{\mathbf{k},\sigma,\alpha}^{\dagger} \right.\nonumber \\
        &\hspace{1.5cm} \left. + \left[P_{\alpha}^{12}(\mathbf{k})\right]^{*}_{(b,r_s),\sigma} \gamma_{-\mathbf{k},\sigma,\alpha}  \right)\\
        d^{\dagger}_{\mathbf{k},\mathbf{r}_s,\alpha} &= \sum_{\sigma=1}^{2N_s} \left( \left[P_{\alpha}^{11}(\mathbf{k})\right]^*_{(d,r_s),\sigma} \gamma^\dagger_{\mathbf{k},\sigma,\alpha} \right.\nonumber \\
        &\hspace{1.5cm} \left. + \left[P_{\alpha}^{12}(\mathbf{k})\right]^*_{(d,r_s),\sigma} \gamma_{-\mathbf{k},\sigma,\alpha}  \right),
    \end{align}
\end{subequations}
where the following notation is used 
\begin{subequations}
    \begin{align}
        (d,r_s) &\equiv r_s \\
        (b,r_s) &\equiv r_s + N_s.
    \end{align}
\end{subequations}
These eigenmodes satisfy 
\begin{align}
    \expval{\gamma^{\dagger}_{\mathbf{k},\sigma,\alpha} \gamma_{\mathbf{q}, \rho,\beta}} &= n_{B}(\omega_{\sigma,\alpha}(\mathbf{k})) \delta_{\mathbf{k},\mathbf{q}} \delta_{\sigma,\rho} \delta_{\alpha, \beta},
\end{align}
where $n_{B}(\omega_{\sigma}(\mathbf{k}))$ is the Bose-Einstein distribution. This directly implies that the Bogoliubov operator annihilates the ground state at $T=0$ (i.e., the ground state is a Bogoliubon/spinon vacuum)
\begin{equation}
    \gamma_{\mathbf{k},\sigma,\alpha} \ket{\psi_{GS}} = 0.
\end{equation}

The corresponding ground-state energy per diamond lattice unit cell is
\begin{align}
E_{\text{G.S.}}=\frac{1}{N_{\text{u.c.}}} \sum_{\mathbf{k},\sigma,\alpha} \left(\omega_{\sigma,\alpha}(\mathbf{k})-\frac{1}{2}\right).
\end{align}

\subsection{\label{Appendix SubSec:Analytical diagonalization of the quadratic Hamiltonians - 0-flux} Diagonalization of the 0-flux Hamiltonian}

For the 0-flux phase, the gauge field background is $\bar{A}_{\mathbf{r}_\alpha,\mathbf{r}_\beta'}=0$ everywhere. The corresponding Hamiltonian matrix is 
\begin{align}
    &H_{\alpha}^{\text{0-flux}}\left(\mathbf{k}\right) \nonumber \\
    &= \frac{J_{\parallel}}{2}\mathds{1}_{4 \times 4} - \frac{\tilde{J}_{\pm}}{4} \mqty(
    M(\mathbf{k}) & 0 & 0 & M(\mathbf{k}) \\
    0 & M(\mathbf{k}) & M(\mathbf{k}) & 0 \\
    0 & M(\mathbf{k}) & M(\mathbf{k}) & 0 \\
    M(\mathbf{k}) & 0 & 0 & M(\mathbf{k})),
\end{align}
where we have defined 
\begin{align}
    M(\mathbf{k}) &=  \sum_{i,j\ne i} e^{i\mathbf{k}\cdot (\mathbf{a}_i - \mathbf{a}_{j})} \nonumber \\
    &= 4 \left(\cos\left(\frac{k_x}{2}\right) \cos\left(\frac{k_y}{2}\right) + \cos\left(\frac{k_x}{2}\right)\cos\left(\frac{k_z}{2}\right) \right. \nonumber \\
    &\quad\quad\quad  \left. + \cos\left(\frac{k_y}{2}\right)\cos\left(\frac{k_z}{2}\right)\right).
\end{align}

By direct diagonalization of the matrix $J \mathcal{H}^{\text{0-flux}}(\mathbf{k})$, one can find that there is a single non-degenerate spinon band of the form 
\begin{align}
    \varepsilon(\mathbf{k}) &= \frac{1}{2} \sqrt{ J_{\parallel} \left( J_{\parallel} - \tilde{J}_{\pm} M(\mathbf{k}) \right)},
\end{align}
as stated in the main text.

Furthermore, the Hamiltonian can be diagonalized by a Bogoliubov transformation of the form 
\begin{align}\label{eq: bogoliubov 0-flux}
    P(\mathbf{k}) = \mqty( 
    \cosh(\theta_{\mathbf{k}}) & 0 & 0 & \sinh(\theta_{\mathbf{k}}) \\
    0 & \cosh(\theta_{\mathbf{k}}) & \sinh(\theta_{\mathbf{k}}) & 0 \\
    0 & \sinh(\theta_{\mathbf{k}}) & \cosh(\theta_{\mathbf{k}}) & 0 \\
    \sinh(\theta_{\mathbf{k}}) & 0 & 0 & \cosh(\theta_{\mathbf{k}}) \\
    ),
\end{align}
where $\theta_{\mathbf{k}}$ satisfies
\begin{align}
    &\tanh\left(2 \theta_{\mathbf{k}}\right) = \frac{ \tilde{J}_{\pm} M(\mathbf{k})}{2 J_{\parallel} - \tilde{J}_{\pm} M(\mathbf{k})} \nonumber \\ 
    &\quad \Longrightarrow \theta_{\mathbf{k}} = \frac{1}{2} \ln\left(\sqrt{\frac{J_{\parallel}}{J_{\parallel}-\tilde{J}_{\pm}M(\mathbf{k})}}\right),
\end{align}
such that 
\begin{subequations}
    \begin{align}
        \cosh\left(2 \theta_{\mathbf{k}}\right) &= \frac{(2J_{\parallel} - \tilde{J}_{\pm} M(\mathbf{k}))}{2 \sqrt{J_{\parallel}\left( J_{\parallel} - \tilde{J}_{\pm} M(\mathbf{k}) \right)}} \nonumber \\
        &= \frac{1}{\varepsilon(\mathbf{k})}\left(\frac{J_{\parallel}}{2} - \frac{\tilde{J}_{\pm}}{4} M(\mathbf{k})) \right) \\
        \sinh\left(2 \theta_{\mathbf{k}}\right) &=   \frac{\tilde{J}_{\pm} M(\mathbf{k})}{2\sqrt{J_{\parallel}\left( J_{\parallel} - \tilde{J}_{\pm} M(\mathbf{k}) \right)}} \nonumber \\
        &= \frac{\tilde{J}_{\pm} M(\mathbf{k})}{4 \varepsilon(\mathbf{k})},
    \end{align}
\end{subequations}
as already established in Ref.~\cite{hao2014bosonic}.

\subsection{\label{Appendix SubSec:Analytical diagonalization of the quadratic Hamiltonians - pi-flux} Diagonalization of the \texorpdfstring{$\bm{\pi}$}{pi}-flux Hamiltonian}

In the $\pi$-flux QSI phase, we use the same gauge fixing as in Ref.~\cite{desrochers2023symmetry, desrochers2023spectroscopic}
\begin{subequations}
\begin{align}
\bar{A}_{\left(r_1, r_2, r_3\right)_A,\left(r_1, r_2, r_3\right)_B} & = 0 \\
\bar{A}_{\left(r_1, r_2, r_3\right)_A,\left(r_1+1, r_2, r_3\right)_B} & = n_1 \pi\left(r_2+r_3\right) \\
\bar{A}_{\left(r_1, r_2, r_3\right)_A,\left(r_1, r_2+1, r_3\right)_B} & = n_1 \pi r_3 \\
\bar{A}_{\left(r_1, r_2, r_3\right)_A,\left(r_1, r_2, r_3+1\right)_B} & = 0.
\end{align}
\end{subequations}
The associated Hamiltonian matrix takes the form
\begin{align}
    &H_{\alpha}^{\pi\text{-flux}} \nonumber \\
    &= \frac{J_{\parallel}}{2} \mathds{1}_{16\cross 16} - \frac{\tilde{J}_{\pm}}{2} 
    \left(\begin{array}{@{}c|c|c|c@{}}
    N_{\alpha}(\mathbf{k}) & 0 & 0 & N_{\alpha}(\mathbf{k})  \\[0.5mm] \hline
    0 & N_{\alpha}(\mathbf{k}) & N_\alpha(\mathbf{k}) & 0 \\[0.5mm] \hline
    0 & N_{\alpha}(\mathbf{k}) & N_{\alpha}(\mathbf{k}) & 0 \\[0.5mm] \hline
    N_{\alpha}(\mathbf{k}) & 0 & 0 & N_{\alpha}(\mathbf{k}) 
    \end{array}\right),
\end{align}
with
\begin{widetext}
\begin{subequations}
    \begin{align}
        N_A(\mathbf{k}) &= \left(
        \begin{array}{cccc}
         C(k_y,k_z) & C(k_x,k_z)-iS(k_x,k_y) & C(k_x,k_y)-iS(k_x,k_z) & -i S(k_y,k_z) \\
         C(k_x,k_z)+iS(k_x,k_y) & - C(k_y,k_z) & -i S(k_y,k_z) & C(k_x,k_y)+iS(k_x,k_z) \\
         C(k_x,k_y)+iS(k_x,k_z) & i S(k_y,k_z) & -C(k_y,k_z) & -C(k_x,k_z)-iS(k_x,k_y)\\
         i S(k_y,k_z) & C(k_x,k_y)-iS(k_x,k_z) & -C(k_x,k_z)+iS(k_x,k_y) & C(k_y,k_z) 
        \end{array}\right) \\
        N_B(\mathbf{k}) &= \left(N_A(\mathbf{k})\right)^* ,
    \end{align}
\end{subequations}
\end{widetext}
where we have used the following notation to simplify the presentation
\begin{subequations}
\begin{align}
    C(k_a, k_b) &= \cos\left(\frac{k_a+k_b}{2}\right) = C(k_b, k_a) \\
    S(k_a, k_b) &= \sin\left(\frac{k_a-k_b}{2}\right) = -S(k_b, k_a).
\end{align}
\end{subequations}

By direct diagonalization of the matrix $J H^{\pi\text{-flux}}(\mathbf{k})$, one can find that there are two non-degenerate spinon bands of the form 
\begin{align}
    \varepsilon_{\pm}(\mathbf{k}) &= \frac{1}{2} \sqrt{J_{\parallel} \left( J_{\parallel} \pm 2 |\tilde{J}_{\pm}| \lambda(\mathbf{k}) \right)}
\end{align}
with 
\begin{align}
    \lambda(\mathbf{k}) =&  \left[3 - \sin\left(k_x\right) \sin\left(k_y\right) - \sin\left(k_x\right)\sin\left(k_z\right)  \right. \nonumber \\
    &\quad\quad \left.- \sin\left(k_y\right)\sin\left(k_z\right)\right]^{1/2},
\end{align}
as presented in the main text.

The Hamiltonian can be diagonalized in two steps. First, we perform a block-diagonal unitary transformation of the form (since the matrix is block diagonal, it is a valid Bogoliubov transformation that respects Eq.~\eqref{eq: paraunitary constraint})
\begin{equation}
    \mathcal{U}_{\alpha}(\mathbf{k}) = \left(\begin{array}{@{}c|c|c|c@{}}
    U_{\alpha}(\mathbf{k}) & 0 & 0 & 0  \\[0.5mm] \hline
    0 & U_{\alpha}(\mathbf{k}) & 0 & 0 \\[0.5mm] \hline
    0 & 0 & U_{\alpha}(\mathbf{k}) & 0 \\[0.5mm] \hline
    0 & 0 & 0 & U_{\alpha}(\mathbf{k})
    \end{array}\right),
\end{equation}
such that the $U_{\alpha}(\mathbf{k})$ matrix diagonalizes $N_{\alpha}(\mathbf{k})$
\begin{equation}
    U^{\dagger}_{\alpha}(\mathbf{k}) N_{\alpha}(\mathbf{k}) U_{\alpha}(\mathbf{k}) = \Lambda_{\alpha}(\mathbf{k}).
\end{equation}
$\Lambda_{\alpha}(\mathbf{k})$ is the diagonal matrix
\begin{equation}
    \Lambda(\mathbf{k}) = \text{diag}\left( \lambda(\mathbf{k}), \lambda(\mathbf{k}), -\lambda(\mathbf{k}), -\lambda(\mathbf{k}) \right). 
\end{equation}
Because $N_{\alpha}(\mathbf{k})$ is an hermitian matrix and we have $N_{A}(\mathbf{k})=(N_{B}(\mathbf{k}))^{*}$, it should be noted that
\begin{equation}
    \Lambda_{A}(\mathbf{k}) = \Lambda_{B}(\mathbf{k}) \equiv \Lambda(\mathbf{k})
\end{equation}
and
\begin{equation}
    U_A(\mathbf{k}) = (U_B(\mathbf{k}))^*.
\end{equation}

Therefore, by acting with $\mathcal{U}_{\alpha}(\mathbf{k})$, the Hamiltonian takes the form 
\begin{align}\label{eq: Hamiltonian pi-flux after block diagonalization}
    &\mathcal{U}^{\dagger}_{\alpha}(\mathbf{k}) H_{\alpha}^{\pi\text{-flux}}(\mathbf{k}) \mathcal{U}_{\alpha}(\mathbf{k}) \nonumber \\
    &= \frac{J_{\parallel}}{2} \mathds{1}_{16\cross 16} - \frac{\tilde{J}_{\pm}}{4} 
    \left(\begin{array}{@{}c|c|c|c@{}}
    \Lambda(\mathbf{k}) & 0 & 0 & \Lambda(\mathbf{k}) \\[0.5mm] \hline
    0 & \Lambda(\mathbf{k}) & \Lambda(\mathbf{k}) & 0 \\[0.5mm] \hline
    0 & \Lambda(\mathbf{k}) & \Lambda(\mathbf{k}) & 0 \\[0.5mm] \hline
    \Lambda(\mathbf{k})  & 0 & 0 & \Lambda(\mathbf{k})
    \end{array}\right),
\end{align}
The $U_A(\mathbf{k})$ matrix can be explicitly expressed as 
\begin{align}
    U_A(\mathbf{k}) = 
    \mqty(
    \vert & \vert & \vert & \vert \\
    u_{A,1}(\mathbf{k})   & u_{A,2}(\mathbf{k}) & u_{A,3}(\mathbf{k}) & u_{A,4}(\mathbf{k})   \\
    \vert & \vert & \vert & \vert),
\end{align}
where 
\begingroup
\allowdisplaybreaks
\begin{widetext}
\begin{subequations}
\begin{align}
    u_{A,1}(\mathbf{k}) &= \frac{1}{{2 \sqrt{\lambda(\mathbf{k})  (-C(k_{y},k_{z})+\lambda(\mathbf{k}) +S(k_{y},k_{z}))}}} \left(
    \begin{array}{c}
    i (C(k_{x},k_{y})-i (C(k_{x},k_{z})-i S(k_{x},k_{y})+S(k_{x},k_{z}))) \\
    -C(k_{y},k_{z})+\lambda(\mathbf{k}) +S(k_{y},k_{z}) \\
    i (-C(k_{y},k_{z})+\lambda(\mathbf{k}) +S(k_{y},k_{z})) \\
    C(k_{x},k_{y})-i (C(k_{x},k_{z})-i S(k_{x},k_{y})+S(k_{x},k_{z})) \\
    \end{array}\right) \\
    u_{A,2}(\mathbf{k}) &= \frac{1}{{2 \sqrt{\lambda(\mathbf{k})  (-C(k_{y},k_{z})+\lambda(\mathbf{k}) -S(k_{y},k_{z}))}}} \left(
    \begin{array}{c}
    -i (C(k_{x},k_{y})+i C(k_{x},k_{z})+S(k_{x},k_{y})-i S(k_{x},k_{z}))\\
    -C(k_{y},k_{z})+\lambda(\mathbf{k}) -S(k_{y},k_{z}) \\
    -i (-C(k_{y},k_{z})+\lambda(\mathbf{k}) -S(k_{y},k_{z})) \\
    C(k_{x},k_{y})+i C(k_{x},k_{z})+S(k_{x},k_{y})-i S(k_{x},k_{z}) \\
    \end{array}\right)\\
    u_{A,3}(\mathbf{k}) &= \frac{1}{{2 \sqrt{\lambda(\mathbf{k})  (C(k_{y},k_{z})+\lambda(\mathbf{k}) -S(k_{y},k_{z}))}}} \left(
    \begin{array}{c}
    i (C(k_{x},k_{y})-i (C(k_{x},k_{z})-i S(k_{x},k_{y})+S(k_{x},k_{z}))) \\
    -C(k_{y},k_{z})-\lambda(\mathbf{k}) +S(k_{y},k_{z}) \\
    - i (C(k_{y},k_{z})+\lambda(\mathbf{k}) -S(k_{y},k_{z})) \\
    C(k_{x},k_{y})-i (C(k_{x},k_{z})-i S(k_{x},k_{y})+S(k_{x},k_{z})) \\
    \end{array}\right)\\
    u_{A,4}(\mathbf{k}) &= \frac{1}{{2 \sqrt{\lambda(\mathbf{k})  (C(k_{y},k_{z})+\lambda(\mathbf{k}) +S(k_{y},k_{z}))}}} \left(
    \begin{array}{c} 
    - i (C(k_{x},k_{y})+i C(k_{x},k_{z})+S(k_{x},k_{y})-i S(k_{x},k_{z})) \\
    -C(k_{y},k_{z})-\lambda(\mathbf{k}) -S(k_{y},k_{z}) \\
    i (C(k_{y},k_{z})+\lambda(\mathbf{k}) +S(k_{y},k_{z}))\\
    C(k_{x},k_{y})+i C(k_{x},k_{z})+S(k_{x},k_{y})-i S(k_{x},k_{z}) \\
    \end{array}\right)
\end{align}
\end{subequations}
Next, the Hamiltonian~\eqref{eq: Hamiltonian pi-flux after block diagonalization} can be fully diagonalized by using a transformation of the form 
\begin{align}
    \mathcal{V}(\mathbf{k}) = \left(\begin{array}{@{}c|c|c|c@{}}
    V_{1}(\mathbf{k}) & 0 & 0 & V_{2}(\mathbf{k})  \\[0.5mm] \hline
    0 & V_{1}(\mathbf{k}) & V_{2}(\mathbf{k}) & 0 \\[0.5mm] \hline
    0 & V_{2}(\mathbf{k}) & V_{1}(\mathbf{k}) & 0 \\[0.5mm] \hline
    V_{2}(\mathbf{k}) & 0 & 0 & V_{1}(\mathbf{k}) 
    \end{array}\right),
\end{align}
where 
\begin{subequations}
    \begin{align}
        V_{1}(\mathbf{k}) &=  
        \left(\begin{array}{@{}c|c@{}}
        \cosh(\theta_{1}(\mathbf{k})) \mathds{1}_{2\times 2} & 0   \\[0.5mm] \hline
        0 & \cosh(\theta_{2}(\mathbf{k})) \mathds{1}_{2\times 2}
        \end{array}\right) \\
        V_{2}(\mathbf{k}) &=  
        \left(\begin{array}{@{}c|c@{}}
        \sinh(\theta_{1}(\mathbf{k})) \mathds{1}_{2\times 2} & 0   \\[0.5mm] \hline
        0 & \sinh(\theta_{2}(\mathbf{k})) \mathds{1}_{2\times 2}
        \end{array}\right).
    \end{align}
\end{subequations}
Acting with this transformation after $\mathcal{U}_{\alpha}(\mathbf{k})$, we get 
\begin{align}
    &\mathcal{V}^{\dagger}(\mathbf{k})\mathcal{U}^{\dagger}_{\alpha}(\mathbf{k}) H_{\alpha}^{\pi\text{-flux}}(\mathbf{k}) \mathcal{U}_{\alpha}(\mathbf{k}) \mathcal{V}(\mathbf{k}) \notag \\
    &= \frac{J_{\parallel}}{2}\left(\begin{array}{@{}c|c|c|c@{}}
    (V_{1}(\mathbf{k}))^2 + (V_{2}(\mathbf{k}))^2 & 0 & 0 & 2 V_{1}(\mathbf{k}) V_{2}(\mathbf{k})  \\[0.5mm] \hline
    0 & (V_{1}(\mathbf{k}))^2 + (V_{2}(\mathbf{k}))^2 & 2 V_{1}(\mathbf{k}) V_{2}(\mathbf{k}) & 0 \\[0.5mm] \hline
    0 & 2 V_{1}(\mathbf{k}) V_{2}(\mathbf{k}) & (V_{1}(\mathbf{k}))^2 + (V_{2}(\mathbf{k}))^2 & 0 \\[0.5mm] \hline
    2 V_{1}(\mathbf{k}) V_{2}(\mathbf{k}) & 0 & 0 & (V_{1}(\mathbf{k}))^2 + (V_{2}(\mathbf{k}))^2
    \end{array}\right) \notag \\
    &\quad\quad - \frac{\tilde{J}_{\pm}}{4} (V_{1}(\mathbf{k}) + V_{2}(\mathbf{k})) \Lambda(\mathbf{k}) (V_{1}(\mathbf{k}) + V_{2}(\mathbf{k})) \left(\begin{array}{@{}c|c|c|c@{}}
    \mathds{1}   & 0 & 0 & \mathds{1}  \\[0.5mm] \hline
    0 & \mathds{1} & \mathds{1} & 0 \\[0.5mm] \hline
    0 & \mathds{1} & \mathds{1} & 0 \\[0.5mm] \hline
    \mathds{1} & 0 & 0 & \mathds{1}
    \end{array}\right).
\end{align}
The system is diagonalized if 
\begin{subequations}
    \begin{align}
        &\tanh\left(2 \theta_{1}(\mathbf{k})\right) = \frac{ \tilde{J}_{\pm} \lambda(\mathbf{k})}{2 J_{\parallel} - \tilde{J}_{\pm} \lambda(\mathbf{k})} \Longrightarrow \theta_{1}(\mathbf{k}) = \frac{1}{2} \ln\left(\sqrt{\frac{J_{\parallel}}{J_{\parallel}-\tilde{J}_{\pm}\lambda(\mathbf{k})}}\right) \\
        &\tanh\left(2 \theta_{2}(\mathbf{k})\right) = \frac{ \tilde{J}_{\pm} \lambda(\mathbf{k})}{2 J_{\parallel} + \tilde{J}_{\pm} \lambda(\mathbf{k})}  \Longrightarrow \theta_{2}(\mathbf{k}) = \frac{1}{2} \ln\left(\sqrt{\frac{J_{\parallel}}{J_{\parallel} + \tilde{J}_{\pm}\lambda(\mathbf{k})}}\right).
    \end{align}
\end{subequations}
In summary, the full Bogoliubov transformation that diagonalizes the $\pi$-flux Hamiltonian is
\begin{align}
    P_{\alpha}(\mathbf{k}) = \mathcal{U}_{\alpha}(\mathbf{k}) \mathcal{V}(\mathbf{k}) &= \left(\begin{array}{@{}c|c|c|c@{}}
    U_{\alpha}(\mathbf{k}) & 0 & 0 & 0  \\[0.5mm] \hline
    0 & U_{\alpha}(\mathbf{k}) & 0 & 0 \\[0.5mm] \hline
    0 & 0 & U_{\alpha}(\mathbf{k}) & 0 \\[0.5mm] \hline
    0 & 0 & 0 & U_{\alpha}(\mathbf{k}) 
    \end{array}\right) \left(\begin{array}{@{}c|c|c|c@{}}
    V_{1}(\mathbf{k}) & 0 & 0 & V_{2}(\mathbf{k})  \\[0.5mm] \hline
    0 & V_{1}(\mathbf{k}) & V_{2}(\mathbf{k}) & 0 \\[0.5mm] \hline
    0 & V_{2}(\mathbf{k}) & V_{1}(\mathbf{k}) & 0 \\[0.5mm] \hline
    V_{2}(\mathbf{k}) & 0 & 0 & V_{1}(\mathbf{k}) 
    \end{array}\right) \notag \\
    &= \left(\begin{array}{@{}c|c|c|c@{}}
    U_{\alpha}(\mathbf{k}) V_{1}(\mathbf{k}) & 0 & 0 & U_{\alpha}(\mathbf{k}) V_{2}(\mathbf{k})  \\[0.5mm] \hline
    0 & U_{\alpha}(\mathbf{k}) V_{1}(\mathbf{k}) & U_{\alpha}(\mathbf{k}) V_{2}(\mathbf{k}) & 0 \\[0.5mm] \hline
    0 & U_{\alpha}(\mathbf{k}) V_{2}(\mathbf{k}) & U_{\alpha}(\mathbf{k}) V_{1}(\mathbf{k}) & 0 \\[0.5mm] \hline
    U_{\alpha}(\mathbf{k}) V_{2}(\mathbf{k}) & 0 & 0 & U_{\alpha}(\mathbf{k}) V_{1}(\mathbf{k}) 
    \end{array}\right),
\end{align}
where 
\begin{subequations}
\begin{align}
    U_{\alpha}(\mathbf{k}) V_{1}(\mathbf{k}) &= 
    \mqty(
    \vert & \vert & \vert & \vert \\
    u_{\alpha,1}(\mathbf{k}) \cosh(\theta_{1}(\mathbf{k}))   & u_{\alpha,2}(\mathbf{k})\cosh(\theta_{1}(\mathbf{k})) & u_{\alpha,3}(\mathbf{k})\cosh(\theta_{2}(\mathbf{k})) & u_{\alpha,4}(\mathbf{k}) \cosh(\theta_{2}(\mathbf{k}))  \\
    \vert & \vert & \vert & \vert) \\
    U_{\alpha}(\mathbf{k}) V_{2}(\mathbf{k}) &= 
    \mqty(
    \vert & \vert & \vert & \vert \\
    u_{\alpha,1}(\mathbf{k}) \sinh(\theta_{1}(\mathbf{k}))   & u_{\alpha,2}(\mathbf{k}) \sinh(\theta_{1}(\mathbf{k})) & u_{\alpha,3}(\mathbf{k})\sinh(\theta_{2}(\mathbf{k})) & u_{\alpha,4}(\mathbf{k}) \sinh(\theta_{2}(\mathbf{k}))  \\
    \vert & \vert & \vert & \vert) .
\end{align}
\end{subequations}
\end{widetext}
\endgroup

\section{\label{Appendix Sec: spin structure factor} Spin structure factor}

\subsection{Generalities}

Let's first compute the dynamical spin structure factor in the sublattice-dependant local spin frame (i.e., without the global transverse projector)
\begin{align}
    \mathcal{S}_{\text{LF}}^{ab}(\mathbf{q},\omega) =& \frac{1}{N_{\text{d.u.c.}}} \sum_{\mathbf{R}_{i}, \mathbf{R}_{j}'} e^{i\mathbf{q}\cdot\left(\mathbf{R}_i - \mathbf{R}_j'\right)} \nonumber \\
    &\quad\quad \times \int dt e^{i\omega t} \expval{\mathrm{S}^{a}_{\mathbf{R}_{i}}(t) \mathrm{S}^{b}_{\mathbf{R}_{j}'}(0)}. 
\end{align}
The matter sector gives access to the transverse components $\mathcal{S}^{+-}$ and $\mathcal{S}^{-+}$. Let us evaluate them in general using the SCEBR. In the following, diamond lattice positions will be denoted by 
\begin{align}
\mathbf{r}_\alpha=\mathbf{r}_{u . c .}+\mathbf{r}_s-\frac{\eta_\alpha}{2} \mathbf{b}_0,
\end{align}
with $\mathbf{r}_{u . c .}$ and $\mathbf{r}_s$ labeling the position of the GMFT Ansatz unit cell and sublattice, respectively. $r_s$ will denote the index (used in constructing the Hamiltonian and Bogoliubov transformation) for the sublattice $\mathbf{r}_s$. The pyrochlore sublattice indices $\mu,\nu=0,1,2,3$ will also be employed.

First, for $\mathcal{S}^{+-}(\mathbf{q},\omega)$ we have
\begingroup
\allowdisplaybreaks
\begin{widetext}
\begin{align}
    &\mathcal{S}^{+-}(\mathbf{q},\omega)\notag \\
    &= \frac{1}{N_{\text{d.u.c.}}} \sum_{\mathbf{R}_{i}, \mathbf{R}_j'} e^{i\mathbf{q}\cdot\left(\mathbf{R}_i - \mathbf{R}_j'\right)} \int dt e^{i\omega t} \expval{\mathrm{S}^{+}_{\mathbf{R}_{i}}(t) \mathrm{S}^{-}_{\mathbf{R}_{j}'}(0)} \notag \\
    &= \frac{1}{N_{\text{d.u.c.}}} \sum_{\mathbf{r}_{A}, \mathbf{r}_A'} \sum_{\mu, \nu} e^{i\mathbf{q}\cdot\left(\mathbf{r}_{A} - \mathbf{r}_{A}' + (\mathbf{b}_{\mu} - \mathbf{b}_{\nu})/2 \right)} \int dt e^{i\omega t} \expval{\mathrm{S}^{+}_{\mathbf{r}_{A}+\mathbf{b}_{\mu}/2}(t) \mathrm{S}^{-}_{\mathbf{r}_{A}'+\mathbf{b}_{\nu}/2}(0)} \notag \\
    &= \frac{1}{N_{\text{d.u.c.}}} \sum_{\mathbf{r}_{A}, \mathbf{r}_A'} \sum_{\mu, \nu} e^{i\mathbf{q}\cdot\left(\mathbf{r}_{A} - \mathbf{r}_{A}' + (\mathbf{b}_{\mu} - \mathbf{b}_{\nu})/2 \right)} \int dt e^{i\omega t} \frac{1}{4} \expval{\Phi^{\dagger}_{\mathbf{r}_A}(t) e^{i \overline{A}_{\mathbf{r}_A,\mathbf{r}_A+\mathbf{b}_\mu}} \Phi_{\mathbf{r}_A+\mathbf{b}_\mu}(t) \Phi^{\dagger}_{\mathbf{r}_A'+\mathbf{b}_\nu}(0) e^{-i \overline{A}_{\mathbf{r}_A',\mathbf{r}_A'+\mathbf{b}_\nu}} \Phi_{\mathbf{r}_A'}(0)  } \notag \\
    &= \frac{1}{4N_{\text{d.u.c.}}(1+\expval{n})^2} \sum_{\mathbf{r}_{A}, \mathbf{r}_A'} \sum_{\mu, \nu} e^{i\mathbf{q}\cdot\left(\mathbf{r}_{A} - \mathbf{r}_{A}' + (\mathbf{b}_{\mu} - \mathbf{b}_{\nu})/2 \right)} e^{i (\overline{A}_{\mathbf{r}_A,\mathbf{r}_A+\mathbf{b}_\mu} - \overline{A}_{\mathbf{r}_A',\mathbf{r}_A'+\mathbf{b}_\nu})}  \notag \\
    &\quad\quad \times \int dt e^{i\omega t} \expval{ \left( b_{\mathbf{r}_A}(t) + d^{\dagger}_{\mathbf{r}_A}(t)  \right) \left( b^{\dagger}_{\mathbf{r}_A+\mathbf{b}_\mu}(t) + d_{\mathbf{r}_A+\mathbf{b}_\mu}(t)  \right)  \left( b_{\mathbf{r}_A'+\mathbf{b}_\nu}(0) + d^{\dagger}_{\mathbf{r}_A'+\mathbf{b}_\nu}(0)  \right) \left( b^{\dagger}_{\mathbf{r}_A'}(0) + d_{\mathbf{r}_A'}(0)  \right)} \notag \\
    &= \frac{1}{4N_{\text{d.u.c.}} N_{\text{u.c.}}^2 (1+\expval{n})^2} \sum_{\mathbf{r}_{\text{u.c.}}, \mathbf{r}_{\text{u.c.}}'} \sum_{\mathbf{r}_{s}, \mathbf{r}_{s}'} \sum_{\mu, \nu} e^{i\mathbf{q}\cdot\left( \left(\mathbf{r}_{\text{u.c.}} + \mathbf{r}_s - \frac{1}{2} \mathbf{b}_0\right)  - \left(\mathbf{r}_{\text{u.c.}}' + \mathbf{r}_s' - \frac{1}{2} \mathbf{b}_0 \right)+ (\mathbf{b}_{\mu} - \mathbf{b}_{\nu})/2 \right)} e^{i \left(\overline{A}_{(\mathbf{r}_s,A),(\mathbf{r}_s+\mathbf{b}_\mu,B)} - \overline{A}_{(\mathbf{r}_s',A),(\mathbf{r}_s'+\mathbf{b}_\nu,B)}\right)}  \notag \\
    &\quad\quad \times \int dt e^{i\omega t} \left\langle\sum_{\mathbf{k}_1}\left( b_{\mathbf{k}_1, \mathbf{r}_s, A}(t)e^{i\mathbf{k}_1\cdot\left( \mathbf{r}_{\text{u.c.}} + \mathbf{r}_s -  \mathbf{b}_0/2 \right)} + d^{\dagger}_{\mathbf{k}_1, \mathbf{r}_s, A}(t) e^{-i\mathbf{k}_1\cdot\left( \mathbf{r}_{\text{u.c.}} + \mathbf{r}_s -  \mathbf{b}_0/2 \right)}  \right) \right. \notag \\
    &\left. \hspace{2.65cm} \sum_{\mathbf{k}_2}\left( b^{\dagger}_{\mathbf{k}_2, \mathbf{r}_s', A}(0) e^{-i\mathbf{k}_2\cdot\left( \mathbf{r}_{\text{u.c.}}' + \mathbf{r}_s' -  \mathbf{b}_0/2 \right)} + d_{\mathbf{k}_2, \mathbf{r}_s', A}(0) e^{i\mathbf{k}_2\cdot\left( \mathbf{r}_{\text{u.c.}}' + \mathbf{r}_s' -  \mathbf{b}_0/2 \right)}  \right) \right\rangle\notag\\
    &\hspace{2.5cm} \left\langle \sum_{\mathbf{k}_3} \left( b^{\dagger}_{\mathbf{k}_3, \mathbf{r}_s+\mathbf{e}_\mu, B}(t) e^{-i\mathbf{k}_3\cdot\left( \mathbf{r}_{\text{u.c.}} + \mathbf{r}_s -  \mathbf{b}_0/2 +\mathbf{b}_\mu \right)} + d_{\mathbf{k}_3, \mathbf{r}_s+\mathbf{e}_\mu, B}(t) e^{i\mathbf{k}_3\cdot\left( \mathbf{r}_{\text{u.c.}} + \mathbf{r}_s -  \mathbf{b}_0/2 + \mathbf{b}_{\mu} \right)}  \right)
    \right. \notag \\
    &\left.  \hspace{2.75cm} \sum_{\mathbf{k}_4} \left( b_{\mathbf{k}_4, \mathbf{r}_s'+\mathbf{e}_\nu, B}(0) e^{i\mathbf{k}_4\cdot\left( \mathbf{r}_{\text{u.c.}}' + \mathbf{r}_s' -  \mathbf{b}_0/2 +\mathbf{b}_\nu \right)} + d^{\dagger}_{\mathbf{k}_4, \mathbf{r}_s'+\mathbf{e}_\nu, B}(0) e^{-i\mathbf{k}_4\cdot\left( \mathbf{r}_{\text{u.c.}}' + \mathbf{r}_s' -  \mathbf{b}_0/2 + \mathbf{b}_{\nu} \right)}  \right) \right\rangle \notag \\
    &= \frac{1}{4N_{\text{d.u.c.}}  (1+\expval{n})^2} \sum_{\substack{\mathbf{k}_1,\mathbf{k}_2,\\ \mathbf{k}_3, \mathbf{k_4}}} \sum_{\mathbf{r}_{s}, \mathbf{r}_{s}'} \sum_{\mu, \nu} e^{i\mathbf{q}\cdot\left( \left( \mathbf{r}_s - \mathbf{r}_s' \right) + (\mathbf{b}_{\mu} - \mathbf{b}_{\nu})/2 \right)} e^{i \left(\overline{A}_{(\mathbf{r}_s,A),(\mathbf{r}_s+\mathbf{b}_\mu,B)} - \overline{A}_{(\mathbf{r}_s',A),(\mathbf{r}_s'+\mathbf{b}_\nu,B)}\right)} \delta_{\mathbf{q},\mathbf{k}_1-\mathbf{k}_3} \delta_{\mathbf{q},\mathbf{k}_2-\mathbf{k}_4}  \notag \\
    &\quad \times \int dt e^{i\omega t}  \left\langle\left( b_{-\mathbf{k}_1, \mathbf{r}_s, A}(t)  + d^{\dagger}_{\mathbf{k}_1, \mathbf{r}_s, A}(t)   \right)   \left( b^{\dagger}_{-\mathbf{k}_2, \mathbf{r}_s', A}(0)  + d_{\mathbf{k}_2, \mathbf{r}_s', A}(0)  \right) \right\rangle e^{-i\left(\mathbf{k}_1\cdot\left( \mathbf{r}_s -  \mathbf{b}_0/2 \right) - \mathbf{k}_2\cdot\left(  \mathbf{r}_s' -  \mathbf{b}_0/2 \right)\right)} \notag\\
    &\quad\times \left\langle \left( b^{\dagger}_{-\mathbf{k}_3, \mathbf{r}_s+\mathbf{e}_\mu, B}(t) + d_{\mathbf{k}_3, \mathbf{r}_s+\mathbf{e}_\mu, B}(t) \right)  \left( b_{-\mathbf{k}_4, \mathbf{r}_s'+\mathbf{e}_\nu, B}(0)  + d^{\dagger}_{\mathbf{k}_4, \mathbf{r}_s'+\mathbf{e}_\nu, B}(0) \right) \right\rangle e^{i\left(-\mathbf{k}_4\cdot\left( \mathbf{r}_s' -  \mathbf{b}_0/2 +\mathbf{b}_\nu \right) +\mathbf{k}_3\cdot\left( \mathbf{r}_s -  \mathbf{b}_0/2 +\mathbf{b}_\mu \right) \right)} \notag 
\end{align}
Each expectation value can be computed separately. For the expectation value on the $A$ sublattice, we have 
\begin{align}
    &\left\langle \left( b_{-\mathbf{k}_1, \mathbf{r}_s, A}(t) + d^{\dagger}_{\mathbf{k}_1, \mathbf{r}_s, A}(t)  \right)  \left( b^{\dagger}_{-\mathbf{k}_2, \mathbf{r}_s', A}(0)  + d_{\mathbf{k}_2, \mathbf{r}_s', A}(0)  \right) \right\rangle \notag \\
    =&  \left\langle \sum_{\sigma} \left( \left[ P_{A}^{11}(\mathbf{k}_1) \right]^{*}_{(d,r_s),\sigma} \gamma_{\mathbf{k}_1,\sigma,A}^{\dagger}(t)  + \left[ P_{A}^{12}(\mathbf{k}_1) \right]^{*}_{(d,r_s),\sigma} \gamma_{-\mathbf{k}_1,\sigma,A}(t) \right.\right. \notag \\
    &\hspace{1cm} \left. + \left[ P_{A}^{11}(-\mathbf{k}_1) \right]_{(b,r_s),\sigma} \gamma_{-\mathbf{k}_1,\sigma,A}(t) + \left[ P_{A}^{12}(-\mathbf{k}_1) \right]_{(b,r_s),\sigma} \gamma_{\mathbf{k}_1,\sigma,A}^{\dagger}(t) \right)  \notag \\
    &\quad\left.  \sum_{\delta} \left( \left[ P_{A}^{11}(\mathbf{k}_2) \right]_{(d,r_s'),\delta} \gamma_{\mathbf{k}_2,\delta,A}(0)  + \left[ P_{A}^{12}(\mathbf{k}_2) \right]_{(d,r_s'),\delta} \gamma_{-\mathbf{k}_2,\delta,A}^{\dagger}(0) \right. \right. \notag \\
    &\hspace{1cm} \left. \left.+ \left[ P_{A}^{11}(-\mathbf{k}_2) \right]_{(b,r_s'),\delta}^* \gamma^{\dagger}_{-\mathbf{k}_2,\delta,A}(0) + \left[ P_{A}^{12}(-\mathbf{k}_2) \right]^*_{(b,r_s'),\delta} \gamma_{\mathbf{k}_2,\delta,A}(0) \right)  \right\rangle \notag \\
    =& \sum_{\sigma,\delta} \left[\left( \left[ P_{A}^{11}(\mathbf{k}_1) \right]^{*}_{(d,r_s),\sigma} + \left[ P_{A}^{12}(-\mathbf{k}_1) \right]_{(b,r_s),\sigma} \right) \left( \left[ P_{A}^{11}(\mathbf{k}_2) \right]_{(d,r_s'),\delta} + \left[ P_{A}^{12}(-\mathbf{k}_2) \right]^*_{(b,r_s'),\delta} \right) \expval{\gamma_{\mathbf{k}_1,\sigma,A}^{\dagger}(t) \gamma_{\mathbf{k}_2,\delta,A}(0)}\right. \notag \\
    &\quad \left.+ \left( \left[ P_{A}^{12}(\mathbf{k}_1) \right]^{*}_{(d,r_s),\sigma} + \left[ P_{A}^{11}(-\mathbf{k}_1) \right]_{(b,r_s),\sigma} \right) \left( \left[ P_{A}^{12}(\mathbf{k}_2) \right]_{(d,r_s'),\delta} + \left[ P_{A}^{11}(-\mathbf{k}_2) \right]_{(b,r_s'),\delta}^* \right) \expval{\gamma_{-\mathbf{k}_1,\sigma,A}(t) \gamma_{-\mathbf{k}_2,\delta,A}^{\dagger}(0)}\right] \notag \\
    =& \delta_{\mathbf{k}_1,\mathbf{k}_2} \sum_{\sigma} \left[\left( \left[ P_{A}^{11}(\mathbf{k}_1) \right]^{*}_{(d,r_s),\sigma} + \left[ P_{A}^{21}(\mathbf{k}_1) \right]^*_{(b,r_s),\sigma} \right) \left( \left[ P_{A}^{11}(\mathbf{k}_1) \right]_{(d,r_s'),\sigma}  + \left[ P_{A}^{21}(\mathbf{k}_1) \right]_{(b,r_s'),\sigma} \right) e^{i \omega_{\sigma,A}(\mathbf{k}_1) t} n_{B}\left(\omega_{\sigma,A}(\mathbf{k}_1)\right) \right. \notag \\
    &\quad\quad\quad \left.+ \left( \left[ P_{A}^{12}(\mathbf{k}_1) \right]^{*}_{(d,r_s),\sigma} + \left[ P_{A}^{22}(\mathbf{k}_1) \right]^*_{(b,r_s),\sigma} \right) \left( \left[ P_{A}^{12}(\mathbf{k}_1) \right]_{(d,r_s'),\sigma} + \left[ P_{A}^{22}(\mathbf{k}_1) \right]_{(b,r_s'),\sigma} \right) e^{-i \omega_{\sigma,A}(\mathbf{k}_1) t} \left( 1 + n_{B}\left(\omega_{\sigma,A}(\mathbf{k}_1)\right) \right) \right] \notag \\
    &= \delta_{\mathbf{k}_1,\mathbf{k}_2} \sum_{\sigma} \left[ f_{A,1}(\mathbf{k}_1,r_s,r_s',\sigma) e^{i \omega_{\sigma,A}(\mathbf{k}_1) t} n_{B}\left(\omega_{\sigma,A}(\mathbf{k}_1)\right)  +  f_{A,2}(\mathbf{k}_1,r_s,r_s',\sigma) e^{-i \omega_{\sigma,A}(\mathbf{k}_1) t} \left( 1 + n_{B}\left(\omega_{\sigma,A}(\mathbf{k}_1)\right)\right) \right].
\end{align}
Similarly, the expectation on the $B$ sublattice is
\begin{align}
    &\left\langle \left( b^{\dagger}_{-\mathbf{k}_3, \mathbf{r}_s+\mathbf{e}_\mu, B}(t) + d_{\mathbf{k}_3, \mathbf{r}_s+\mathbf{e}_\mu, B}(t) \right)  \left( b_{-\mathbf{k}_4, \mathbf{r}_s'+\mathbf{e}_\nu, B}(0)  + d^{\dagger}_{\mathbf{k}_4, \mathbf{r}_s'+\mathbf{e}_\nu, B}(0) \right) \right\rangle \nonumber \\
    =&  \left\langle\sum_{\sigma} \left( \left[ P_{B}^{11}(\mathbf{k}_3) \right]_{(d,\mathbf{r}_s+\mathbf{e}_\mu),\sigma} \gamma_{\mathbf{k}_3,\sigma,B}(t)  + \left[ P_{B}^{12}(\mathbf{k}_3) \right]_{(d,\mathbf{r}_s+\mathbf{e}_\mu),\sigma} \gamma_{-\mathbf{k}_3,\sigma,B}^{\dagger}(t) \right.\right.  \notag \\
    &\hspace{1cm} \left.+ \left[ P_{B}^{11}(-\mathbf{k}_3) \right]_{(b,\mathbf{r}_s+\mathbf{e}_\mu),\sigma}^* \gamma^{\dagger}_{-\mathbf{k}_3,\sigma,B}(t) + \left[ P_{B}^{12}(-\mathbf{k}_3) \right]^*_{(b,\mathbf{r}_s+\mathbf{e}_\mu),\sigma} \gamma_{\mathbf{k}_3,\sigma,B}(t) \right) \notag \\
    &\quad \sum_{\delta} \left( \left[ P_{B}^{11}(\mathbf{k}_4) \right]^{*}_{(d,\mathbf{r}_s'+\mathbf{e}_\nu),\delta} \gamma_{\mathbf{k}_4,\delta,B}^{\dagger}(0)  + \left[ P_{B}^{12}(\mathbf{k}_4) \right]^{*}_{(d,\mathbf{r}_s'+\mathbf{e}_\nu),\delta} \gamma_{-\mathbf{k}_4,\delta,B}(0) \right. \notag \\
    &\hspace{1cm} \left.\left. + \left[ P_{B}^{11}(-\mathbf{k}_4) \right]_{(b,\mathbf{r}_s'+\mathbf{e}_\nu),\delta} \gamma_{-\mathbf{k}_4,\delta,B}(0) + \left[ P_{B}^{12}(-\mathbf{k}_4) \right]_{(b,\mathbf{r}_s'+\mathbf{e}_\nu),\delta} \gamma_{\mathbf{k}_4,\delta,B}^{\dagger}(0) \right)\right\rangle  \notag \\
    &= \delta_{\mathbf{k}_3,\mathbf{k}_4} \sum_{\sigma} \left[ \left( \left[ P_{B}^{12}(\mathbf{k}_3) \right]_{(d,\mathbf{r}_s+\mathbf{e}_\mu),\sigma} + \left[ P_{B}^{11}(-\mathbf{k}_3) \right]_{(b,\mathbf{r}_s+\mathbf{e}_\mu),\sigma}^* \right) \right. \notag \\
    &\hspace{2cm} \times \left( \left[ P_{B}^{12}(\mathbf{k}_3) \right]^{*}_{(d,\mathbf{r}_s'+\mathbf{e}_\nu),\sigma} + \left[ P_{B}^{11}(-\mathbf{k}_3) \right]_{(b,\mathbf{r}_s'+\mathbf{e}_\nu),\sigma}\right) e^{i \omega_{\sigma, A}\left(\mathbf{k}_3\right) t} n_B\left(\omega_{\sigma, B}\left(\mathbf{k}_3\right)\right)\notag \\
    &\quad\quad  +  \left( \left[ P_{B}^{11}(\mathbf{k}_3) \right]_{(d,\mathbf{r}_s+\mathbf{e}_\mu),\sigma} + \left[ P_{B}^{12}(-\mathbf{k}_3) \right]^*_{(b,\mathbf{r}_s+\mathbf{e}_\mu),\sigma} \right) \notag \\
    &\hspace{2cm} \left. \times \left( \left[ P_{B}^{11}(\mathbf{k}_3) \right]^{*}_{(d,\mathbf{r}_s'+\mathbf{e}_\nu),\sigma} +  \left[ P_{B}^{12}(-\mathbf{k}_3) \right]_{(b,\mathbf{r}_s'+\mathbf{e}_\nu),\sigma}\right) e^{-i \omega_{\sigma, A}\left(\mathbf{k}_3\right) t}\left(1+n_B\left(\omega_{\sigma, B}\left(\mathbf{k}_3\right)\right)\right) \right] \notag \\
    &= \delta_{\mathbf{k}_3,\mathbf{k}_4} \sum_{\sigma} \left[ f_{B,1}(\mathbf{k}_3,r_s,r_s',\mu,\nu,\sigma) e^{i \omega_{\sigma,B}(\mathbf{k}_3) t} n_{B}\left(\omega_{\sigma,B}(\mathbf{k}_3)\right)  +  f_{B,2}(\mathbf{k}_3,r_s,r_s',\mu,\nu,\sigma) e^{-i \omega_{\sigma,B}(\mathbf{k}_3) t} \left( 1 + n_{B}\left(\omega_{\sigma,B}(\mathbf{k}_3)\right)\right) \right].
\end{align}
In both derivations, we have introduced the quantities
\begin{subequations}
\begin{align}
        f_{A,1}(\mathbf{k}_1,r_s,r_s',\sigma) &= \left(\left[ P_{A}^{11}(\mathbf{k}_1) \right]^{*}_{(d,r_s),\sigma} + \left[ P_{A}^{21}(\mathbf{k}_1) \right]^*_{(b,r_s),\sigma}\right) \left(\left[ P_{A}^{11}(\mathbf{k}_1) \right]_{(d,r_s'),\sigma}  + \left[ P_{A}^{21}(\mathbf{k}_1) \right]_{(b,r_s'),\sigma}\right) \\
        f_{A,2}(\mathbf{k}_1,r_s,r_s',\sigma) &= \left( \left[ P_{A}^{12}(\mathbf{k}_1) \right]^{*}_{(d,r_s),\sigma} + \left[ P_{A}^{22}(\mathbf{k}_1) \right]^*_{(b,r_s),\sigma} \right) \left( \left[ P_{A}^{12}(\mathbf{k}_1) \right]_{(d,r_s'),\sigma} + \left[ P_{A}^{22}(\mathbf{k}_1) \right]_{(b,r_s'),\sigma} \right) \\
        f_{B,1}(\mathbf{k}_3,r_s,r_s',\mu,\nu,\sigma) &= \left( \left[ P_{B}^{12}(\mathbf{k}_3) \right]_{(d,\mathbf{r}_s+\mathbf{e}_\mu),\sigma} + \left[ P_{B}^{22}(\mathbf{k}_3) \right]_{(b,\mathbf{r}_s+\mathbf{e}_\mu),\sigma} \right) \left( \left[ P_{B}^{12}(\mathbf{k}_3) \right]^{*}_{(d,\mathbf{r}_s'+\mathbf{e}_\nu),\sigma} + \left[ P_{B}^{22}(\mathbf{k}_3) \right]^*_{(b,\mathbf{r}_s'+\mathbf{e}_\nu),\sigma}\right) \\
        f_{B,2}(\mathbf{k}_3,r_s,r_s',\mu,\nu,\sigma) &= \left( \left[ P_{B}^{11}(\mathbf{k}_3) \right]_{(d,\mathbf{r}_s+\mathbf{e}_\mu),\sigma} + \left[ P_{B}^{21}(\mathbf{k}_3) \right]_{(b,\mathbf{r}_s+\mathbf{e}_\mu),\sigma} \right) \left( \left[ P_{B}^{11}(\mathbf{k}_3) \right]^{*}_{(d,\mathbf{r}_s'+\mathbf{e}_\nu),\sigma} +  \left[ P_{B}^{21}(\mathbf{k}_3) \right]^*_{(b,\mathbf{r}_s'+\mathbf{e}_\nu),\sigma}\right).
\end{align}
\end{subequations}
Putting everything back together, we have
\begin{align}
    &\mathcal{S}^{+-}(\mathbf{q},\omega)\notag \\
    &= \sum_{\substack{\mathbf{k}_1,\mathbf{k}_2,\\ \mathbf{k}_3, \mathbf{k_4}}} \sum_{\mathbf{r}_{s}, \mathbf{r}_{s}'} \sum_{\mu, \nu} \int dt e^{i\omega t} e^{i\mathbf{q}\cdot\left( \left( \mathbf{r}_s - \mathbf{r}_s' \right) + (\mathbf{b}_{\mu} - \mathbf{b}_{\nu})/2 \right)} e^{i \left(\overline{A}_{(\mathbf{r}_s,A),(\mathbf{r}_s+\mathbf{b}_\mu,B)} - \overline{A}_{(\mathbf{r}_s',A),(\mathbf{r}_s'+\mathbf{b}_\nu,B)}\right)} \frac{\delta_{\mathbf{q},\mathbf{k}_1-\mathbf{k}_3} \delta_{\mathbf{q},\mathbf{k}_2-\mathbf{k}_4} \delta_{\mathbf{k}_1,\mathbf{k}_2} \delta_{\mathbf{k}_3,\mathbf{k}_4}}{4N_{\text{d.u.c.}}  (1+\expval{n})^2}  \notag \\
    & \times  \sum_{\sigma,\delta} \left[ f_{A,1}(\mathbf{k}_1,r_s,r_s',\sigma) e^{i \omega_{\sigma,A}(\mathbf{k}_1) t} n_{B}\left(\omega_{\sigma,A}(\mathbf{k}_1)\right) \right. \nonumber \\
    &\quad\quad \left. +  f_{A,2}(\mathbf{k}_1,r_s,r_s',\sigma) e^{-i \omega_{\sigma,A}(\mathbf{k}_1) t} \left( 1 + n_{B}\left(\omega_{\sigma,A}(\mathbf{k}_1)\right)\right) \right] e^{-i\left(\mathbf{k}_1\cdot\left( \mathbf{r}_s -  \mathbf{b}_0/2 \right) - \mathbf{k}_2\cdot\left(  \mathbf{r}_s' -  \mathbf{b}_0/2 \right)\right)} \notag \\
    & \times  \left[ f_{B,1}(\mathbf{k}_3,r_s,r_s',\mu,\nu,\delta) e^{i \omega_{\delta,B}(\mathbf{k}_3) t} n_{B}\left(\omega_{\delta,B}(\mathbf{k}_3)\right)  \right.\\
    &\quad\quad \left.+  f_{B,2}(\mathbf{k}_3,r_s,r_s',\mu,\nu,\delta) e^{-i \omega_{\delta,B}(\mathbf{k}_3) t} \left( 1 + n_{B}\left(\omega_{\delta,B}(\mathbf{k}_3)\right)\right) \right] e^{i\left(-\mathbf{k}_4\cdot\left( \mathbf{r}_s' -  \mathbf{b}_0/2 +\mathbf{b}_\nu \right) +\mathbf{k}_3\cdot\left( \mathbf{r}_s -  \mathbf{b}_0/2 +\mathbf{b}_\mu \right) \right)} \notag \\
    &= \frac{1}{4N_{\text{d.u.c.}}  (1+\expval{n})^2} \sum_{\mathbf{k}_1} \sum_{\mathbf{r}_{s}, \mathbf{r}_{s}'} \sum_{\mu, \nu} e^{i\mathbf{k}_1\cdot\left( \mathbf{b}_{\mu} - \mathbf{b}_{\nu} \right)} e^{-i\mathbf{q}\cdot\left( \mathbf{b}_{\mu} - \mathbf{b}_{\nu} \right)/2} e^{i \left(\overline{A}_{(\mathbf{r}_s,A),(\mathbf{r}_s+\mathbf{b}_\mu,B)} - \overline{A}_{(\mathbf{r}_s',A),(\mathbf{r}_s'+\mathbf{b}_\nu,B)}\right)} \notag \\
    & \quad \times \sum_{\sigma,\delta} \left[ f_{A,1}(\mathbf{k}_1,r_s,r_s',\sigma)   f_{B,1}(\mathbf{k}_1 - \mathbf{q},r_s,r_s',\mu,\nu,\delta) n_{B}\left(\omega_{\sigma,A}(\mathbf{k}_1)\right) n_{B}\left(\omega_{\delta,B}(\mathbf{k}_1 - \mathbf{q})\right)  \delta\left(\omega + \omega_{\sigma,A}(\mathbf{k}_1)+ \omega_{\delta,B}(\mathbf{k}_1 - \mathbf{q})\right)  \right.  \notag \\
    & \quad + f_{A,1}(\mathbf{k}_1,r_s,r_s',\sigma) f_{B,2}(\mathbf{k}_1 - \mathbf{q},r_s,r_s',\mu,\nu,\delta)   n_{B}\left(\omega_{\sigma,A}(\mathbf{k}_1)\right)\left( 1 + n_{B}\left(\omega_{\delta,B}(\mathbf{k}_1 - \mathbf{q})\right)\right) \delta\left(\omega + \omega_{\sigma,A}(\mathbf{k}_1)  -  \omega_{\delta,B}(\mathbf{k}_1 - \mathbf{q})\right) \notag \\
    & \quad + f_{A,2}(\mathbf{k}_1,r_s,r_s',\sigma)  f_{B,1}(\mathbf{k}_1 - \mathbf{q},r_s,r_s',\mu,\nu,\delta) \left( 1 + n_{B}\left(\omega_{\sigma,A}(\mathbf{k}_1)\right)\right) n_{B}\left(\omega_{\delta,B}(\mathbf{k}_1 - \mathbf{q})\right) \delta\left(\omega -\omega_{\sigma,A}(\mathbf{k}_1)  + \omega_{\delta,B}(\mathbf{k}_1 - \mathbf{q})\right) \notag \\
    & \quad \left.+ f_{A,2}(\mathbf{k}_1,r_s,r_s',\sigma) f_{B,2}(\mathbf{k}_1 - \mathbf{q},r_s,r_s',\mu,\nu,\delta) \left( 1 + n_{B}\left(\omega_{\sigma,A}(\mathbf{k}_1)\right)\right)  \left( 1 + n_{B}\left(\omega_{\delta,B}(\mathbf{k}_1 - \mathbf{q})\right)\right) \delta\left(\omega-\omega_{\sigma,A}(\mathbf{k}_1)  - \omega_{\delta,B}(\mathbf{k}_1 - \mathbf{q})\right) \right]. \notag \\
\end{align}

To evaluate $\mathcal{S}^{-+}(\mathbf{q},\omega)$, we proceed in a completely analogous manner to find
\begin{align}
    &\mathcal{S}^{-+}(\mathbf{q},\omega)\notag \\
    &= \frac{1}{4N_{\text{d.u.c.}}  (1+\expval{n})^2} \sum_{\mathbf{k}} \sum_{\mathbf{r}_{s}, \mathbf{r}_{s}'} \sum_{\mu, \nu}  e^{-i \mathbf{k}\cdot\left( \mathbf{b}_\mu -   \mathbf{b}_\nu \right) } e^{i\mathbf{q}\cdot\left(  \mathbf{b}_{\mu} - \mathbf{b}_{\nu} \right)/2} e^{-i \left(\overline{A}_{(\mathbf{r}_s,A),(\mathbf{r}_s+\mathbf{b}_\mu,B)} - \overline{A}_{(\mathbf{r}_s',A),(\mathbf{r}_s'+\mathbf{b}_\nu,B)}\right)} \notag \\
    & \quad \times  \sum_{\sigma,\delta} \left[ f^*_{A,1}(\mathbf{k}-\mathbf{q},r_s,r_s',\sigma) f^*_{B,1}( \mathbf{k},r_s,r_s',\mu,\nu,\delta)  \left(1+n_{B}\left(\omega_{\sigma,A}(\mathbf{k}-\mathbf{q})\right)\right) \left(1+n_{B}\left(\omega_{\delta,B}( \mathbf{k})\right)\right) \delta\left(\omega - \omega_{\sigma,A}(\mathbf{k}-\mathbf{q}) - \omega_{\delta,B}(\mathbf{k})\right)  \right. \notag \\
    &\hspace{1.1cm} + f^*_{A,1}(\mathbf{k}-\mathbf{q},,r_s,r_s',\sigma) f^*_{B,2}(\mathbf{k},r_s,r_s',\mu,\nu,\delta)  \left(1+n_{B}\left(\omega_{\sigma,A}(\mathbf{k}-\mathbf{q})\right)\right) n_{B}\left(\omega_{\delta,B}(\mathbf{k})\right)  \delta(\omega - \omega_{\sigma,A}(\mathbf{k}-\mathbf{q}) + \omega_{\delta,B}( \mathbf{k}))  \notag \\
    &\hspace{1.1cm} + f^*_{A,2}(\mathbf{k}-\mathbf{q},,r_s,r_s',\sigma) f^*_{B,1}( \mathbf{k},r_s,r_s',\mu,\nu,\delta)  n_{B}\left(\omega_{\sigma,A}(\mathbf{k}-\mathbf{q})\right) \left( 1+  n_{B}\left(\omega_{\delta,B}(\mathbf{k})\right) \right) \delta(\omega + \omega_{\sigma,A}(\mathbf{k}-\mathbf{q}) - \omega_{\delta,B}(\mathbf{k}))  \notag \\
    &\hspace{1.1cm} \left. + f^*_{A,2}(\mathbf{k}-\mathbf{q},,r_s,r_s',\sigma) f^*_{B,2}(\mathbf{k},r_s,r_s',\mu,\nu,\delta)  n_{B}\left(\omega_{\sigma,A}(\mathbf{k}-\mathbf{q})\right)  n_{B}\left(\omega_{\delta,B}(\mathbf{k})\right) \delta(\omega + \omega_{\sigma,A}(\mathbf{k}-\mathbf{q}) + \omega_{\delta,B}( \mathbf{k})) \right]. 
\end{align}
\end{widetext}
\endgroup

\subsection{Spin structure factor for 0-flux QSI}

In the 0-flux phase, $\overline{A}=0$ everywhere and there is a single sublattice
\begin{equation}
    \mathbf{r}_s \in \{(0,0,0)\} \Longrightarrow r_s \in \{0\},
\end{equation}
such that $f_{A,1}(\mathbf{k},r_s,r_s',\sigma)=f_{A,1}(\mathbf{k},\sigma)$ and $f_{B,1}(\mathbf{k},r_s,r_s',\mu,\nu,\sigma)=f_{B,1}(\mathbf{k},\sigma)$. Furthermore, the Bogoliubov transformation that diagonalizes the system is given in Eq.~\eqref{eq: bogoliubov 0-flux}. We then have
\begin{subequations}
    \begin{align}
        P^{11}(\mathbf{k}) &= P^{22}(\mathbf{k})= \mqty(\cosh \left(\theta_{\mathbf{k}}\right) & 0 \\ 0 & \cosh \left(\theta_{\mathbf{k}}\right)) \\
        P^{12}(\mathbf{k}) &= P^{21}(\mathbf{k})=  \mqty(0 & \sinh\left(\theta_{\mathbf{k}}\right) \\ \sinh\left(\theta_{\mathbf{k}}\right) & 0).
    \end{align}
\end{subequations}
Using this analytical form of the Bogoliubov transformation, the $f$ components are 
\begin{subequations}
    \begin{align}
        f_{A,1}(\mathbf{k},\sigma) &= f_{B,2}(\mathbf{k},\sigma) = \mqty( e^{2\theta_{\mathbf{k}}} \\ 0)_{\sigma}  \\
        f_{A,2}(\mathbf{k},\sigma) &= f_{B,1}(\mathbf{k},\sigma) = \mqty( 0 \\ e^{2\theta_{\mathbf{k}}} )_{\sigma}.
    \end{align}
\end{subequations}
The factors in the evaluation of the dynamical spin structure factor are then
\begin{align}
    &\sum_{\sigma,\delta} f_{A, 1}\left(\mathbf{k}_1, \sigma\right) f_{B, 1}\left(\mathbf{k}_1-\mathbf{q}, \delta\right) \nonumber \\
    =& \sum_{\sigma,\delta} f_{A, 1}\left(\mathbf{k}_1, \sigma\right) f_{B, 2}\left(\mathbf{k}_1-\mathbf{q}, \delta\right) \nonumber \\
    =& \sum_{\sigma,\delta} f_{A, 2}\left(\mathbf{k}_1, \sigma\right) f_{B, 1}\left(\mathbf{k}_1-\mathbf{q}, \delta\right) \nonumber \\
    =& \sum_{\sigma,\delta} f_{A, 2}\left(\mathbf{k}_1, \sigma\right) f_{B, 2}\left(\mathbf{k}_1-\mathbf{q}, \delta\right)\nonumber \\
    =& \frac{J_{xx}}{\sqrt{(J_{xx}-\tilde{J}_{\pm}M(\mathbf{k}_1))(J_{xx}-\tilde{J}_{\pm}M(\mathbf{k}_1 - \mathbf{q}))}}.
\end{align}
\begingroup
\allowdisplaybreaks
\begin{widetext}
We thus have the following final form for the dynamical structure factor in the 0-flux phase
\begin{align}
    &\mathcal{S}^{+-}(\mathbf{q},\omega)= \mathcal{S}^{-+}(\mathbf{q},\omega) \notag \\
    &= \frac{1}{4N_{\text{d.u.c.}}  (1+\expval{n})^2} \sum_{\mu, \nu} e^{-i\mathbf{q}\cdot\left( \mathbf{b}_{\mu} - \mathbf{b}_{\nu} \right)/2} \sum_{\mathbf{k}} e^{i\mathbf{k}\cdot\left( \mathbf{b}_{\mu} - \mathbf{b}_{\nu} \right)}  \frac{J_{xx}}{\sqrt{(J_{xx}-\tilde{J}_{\pm}M(\mathbf{k}))(J_{xx}-\tilde{J}_{\pm}M(\mathbf{k} - \mathbf{q}))}} \notag \\
    & \quad \times \left[ n_{B}\left(\omega(\mathbf{k})\right) n_{B}\left(\omega(\mathbf{k} - \mathbf{q})\right)  \delta\left(\omega + \omega(\mathbf{k})+ \omega(\mathbf{k} - \mathbf{q})\right)  \right. \notag \\  
    &\quad\quad + n_{B}\left(\omega(\mathbf{k})\right)\left( 1 + n_{B}\left(\omega(\mathbf{k} - \mathbf{q})\right)\right) \delta\left(\omega + \omega(\mathbf{k})  -  \omega(\mathbf{k} - \mathbf{q})\right) \notag \\
    & \quad\quad + \left( 1 + n_{B}\left(\omega(\mathbf{k})\right)\right) n_{B}\left(\omega(\mathbf{k} - \mathbf{q})\right) \delta\left(\omega -\omega(\mathbf{k})  + \omega(\mathbf{k} - \mathbf{q})\right) \notag \\
    & \quad\quad \left.+ \left( 1 + n_{B}\left(\omega(\mathbf{k})\right)\right)  \left( 1 + n_{B}\left(\omega(\mathbf{k} - \mathbf{q})\right)\right) \delta\left(\omega-\omega(\mathbf{k})  - \omega(\mathbf{k} - \mathbf{q})\right) \right]  \notag \\ 
    &= \frac{1}{N_{\text{d.u.c.}} (1+\expval{n})^2} \sum_{\mathbf{k}} \frac{J_{xx}}{\sqrt{(J_{xx}-\tilde{J}_{\pm}M(\mathbf{k}))(J_{xx}-\tilde{J}_{\pm}M(\mathbf{k} - \mathbf{q}))}}  \notag \\
    & \quad \times  \left( 1 + \cos\left(\frac{k_x}{2}- \frac{q_x}{4}\right)\cos\left(\frac{k_y}{2}-\frac{q_y}{4}\right) + \cos\left(\frac{k_x}{2}-\frac{q_x}{4}\right)\cos\left(\frac{k_z}{2}-\frac{q_z}{4}\right)  + \cos\left(\frac{k_y}{2}-\frac{q_y}{4}\right)\cos\left(\frac{k_z}{2}-\frac{q_z}{4}\right) \right) \notag \\
    &\quad\times \left[ n_{B}\left(\omega(\mathbf{k})\right) n_{B}\left(\omega(\mathbf{k} - \mathbf{q})\right)  \delta\left(\omega + \omega(\mathbf{k})+ \omega(\mathbf{k} - \mathbf{q})\right)  \right. \notag \\  
    & \quad\quad + n_{B}\left(\omega(\mathbf{k})\right)\left( 1 + n_{B}\left(\omega(\mathbf{k} - \mathbf{q})\right)\right) \delta\left(\omega + \omega(\mathbf{k})  -  \omega(\mathbf{k} - \mathbf{q})\right) \notag \\
    & \quad\quad + \left( 1 + n_{B}\left(\omega(\mathbf{k})\right)\right) n_{B}\left(\omega(\mathbf{k} - \mathbf{q})\right) \delta\left(\omega -\omega(\mathbf{k})  + \omega(\mathbf{k} - \mathbf{q})\right) \notag \\
    & \quad\quad \left.+ \left( 1 + n_{B}\left(\omega(\mathbf{k})\right)\right)  \left( 1 + n_{B}\left(\omega(\mathbf{k} - \mathbf{q})\right)\right) \delta\left(\omega-\omega(\mathbf{k})  - \omega(\mathbf{k} - \mathbf{q})\right) \right].  
\end{align}
\end{widetext}
\endgroup

%

\end{document}